\documentclass[showpacs,preprint,amsmath,amssymb]{revtex4}
\usepackage{graphicx}
\usepackage{dcolumn}
\usepackage{bm}

\def\be{\begin{equation}}
\def\ee{\end{equation}}
\def\bea{\begin{eqnarray}}
\def\eea{\end{eqnarray}}
\def\line{\hbox to \hsize}

\def\vev #1{{\langle #1\rangle}}
\def\1{\mbox{\bf 1}}

\begin{document}

\title{Explicit monodromy of Moore-Read wave functions on a torus}

\author{Suk Bum Chung}

\affiliation{University of Illinois, Department of Physics\\ 1110 W.
Green St.\\
Urbana, IL 61801 USA\\E-mail: sukchung@uiuc.edu}

\author{Michael Stone}

\affiliation{University of Illinois, Department of Physics\\ 1110 W.
Green St.\\
Urbana, IL 61801 USA\\E-mail: m-stone5@uiuc.edu}

\date{\today}

\begin{abstract}
We construct  the wave functions for the Moore-Read $\nu = 5/2$
quantum Hall state on a torus in the presence of  two quasiholes.
These explicit  wave functions allow us to compute  the  monodromy
matrix that describes the effect of  quasihole motion on the space
of degenerate ground states. The result agrees with the recent 
discussion by Oshikawa {\it et al.} (Annals of Physics {\bf 322} 
1477). Our calculation provides a conformal field theory explanation 
of why certain transitions between ground states are forbidden. It 
is because taking a quasihole around a generator of the torus can 
change the fusion channel of the two quasiholes, and this requires 
a change of parity of the  electron number  in some of the  ground 
states.
\end{abstract}


\maketitle

\section{Introduction}

The degeneracy, first noted in  numerical work by Yoshioka {\it et
al.}\ and W.~P.~Su \cite{gs_deg}, of the  fractional quantum Hall
ground state on torus  is  an essential feature of such  states. In
its absence, Laughlin's general gauge argument would require the
Hall conductance to be an integer. Explicit wave functions for
Laughlin states on torus were written down  by Haldane and Rezayi
\cite{HR_deg_wf}, who identified the degeneracy as arising from
translations of the  center of mass of the incompressible Hall
fluid.  Later, Wen and Niu realized \cite{wen-niu} that the
degeneracy persists even in the absence of translation symmetry, and
that the degree of degeneracy was sensitive to the global topology
of the space in which the Hall fluid resides. This observation lead
them to  introduce the  notion of {\it topological order\/} as a
characterization of  the strongly-correlated ground state. Moore and
Read \cite{moore-read} revealed the more general structure of
topological order by pointing out its connection to rational
conformal field theory. Since topological order is distinguished by
the quantum numbers of ground state and excitations, a change of
topological order requires a quantum phase transition corresponding
to a substantial rearrangement of the many-body Hilbert space. Such
phase transitions are not usually associated with symmetry breaking,
and the resulting topological phases posses no conventional order
parameters. They therefore  fall outside the conventional Landau
theory of phase transitions.

It is possible for a quantum  fluid  to have topologically
degenerate states even when it lives on the plane. The degree of
degeneracy then depends on the number and type of vortex-like
defects in the fluid. Braiding these defects produces transitions
between the topologically protected degenerate states, and it has
been suggested that such manipulations may be exploited  for quantum
computation \cite{usual-suspects}. One such topological phase,  the
Moore-Read Pfaffian state,  is the likely candidate for the observed
$\nu = 5/2$ quantum Hall state \cite{moore-read}. The  Pfaffian wave
function is easy to write down on the plane. When the fluid is
placed on a torus, however,  the wave function becomes more
complicated. Counting the number of degenerate states is harder than
counting the number states for a Laughlin fractional quantum Hall
state, and the answer depends on whether the number of electrons in
the system is odd or even. Nonetheless, the ground state wave
functions have been constructed \cite{greiter-wen,
read-rezayi_paired, read-green} ---  although  some issues still
remain. Using general principles that do not require knowledge of
the explicit Pfaffian wave functions,  and by building on what is
understood for Abelian quasiparticles \cite{masaki-senthil_op},
Oshikawa {\it et al.\/} \cite{masaki} have identified the
topological quasi-particle operations that transform one Moore-Read
ground state to another.

In this paper we  investigate  how the  topological operations   of
Oshikawa {\it et al.\/}'s   affect the    many-electron wave
functions. To do this,  we first construct the Moore-Read Pfaffian
states on a torus in the presence of  two quasihole excitations. We
then explicitly exhibit their transformation properties under
topological operations on  the quasihole positions. In this
process, we provide a  conformal field theory explanation of why the
the monodromy matrix acting on  the space of the degenerate ground
states should be block-diagonal with respect to the parity of the
spin structure. It is because the spin structure dictates both the
fusion channel of two quasiholes and the parity of the number of
electrons.

\section{Laughlin states}

Before tackling the Pfaffian state, it helps to  recall the
properties of the simpler Laughlin states on a torus
\cite{HR_deg_wf,read-rezayi_paired,stone_book}. A torus  can be
regarded  as a rectangle with sides $L_x$ and $L_y$ periodic
boundary conditions. In the Landau gauge, with ${\bf A} = -By\,{\bf
{\hat x}}$,  the periodic boundary conditions on the wave function
are twisted by  a  gauge transformation that is necessary to prepare
a particle that leaves the top edge  of the rectangle for its
reappearance  at the bottom  edge.
 For a single electron we have
\begin{equation}
\psi(x + L_x, y) = \psi(x,y) \,\,\,\,\,\,\,\,\,\,\,\,\,\,\,\,\,
\psi(x,y+L_y) = \exp (-i L_y x / l^2) \psi(x,y), \label{EQ:BC1}
\end{equation}
where $l = \sqrt{\hbar / e B}$ is the magnetic length of the system.
This twisted boundary condition must continue to hold, particle  by
particle, for the many-particle wave function. In our  Landau gauge,
the  lowest Landau level  many-particle wave function takes the form
\begin{equation}
\Psi({\bf r}_1, {\bf r}_2,  \ldots, {\bf r}_{N_e}) = \exp(-\sum_i
y_i^2/2 l^2) f(z_1, z_2, \ldots, z_{N_e}), \label{EQ:LLL}
\end{equation}
where $z_i = (x_i + i y_i)/L_x$ and $f$ is a  holomorphic function
in each of the $z_i$. Combining Eq.~(\ref{EQ:BC1}) and
Eq.~(\ref{EQ:LLL}) shows that $f$ must  satisfy the following
quasi-periodicity  conditions:
\begin{eqnarray}
f(z_1, \ldots, z_i+1, \ldots ) &=& f(z_1, \ldots, z_i, \ldots
),\nonumber\\
f(z_1, \ldots, z_i+\tau, \ldots ) &=& \exp[-i \pi N_s (2z_i +
\tau)]. f(z_1, \ldots, z_i, \ldots ), \label{EQ:BC2}
\end{eqnarray}
Here $\tau = i L_y/L_x$ and $N_s =  L_x L_y/2\pi l^2$ is the number
of flux quanta passing through the torus. (These conditions on $f$
preserve   their form  for a torus which is a periodic
parallelogram, rather than a rectangle. In this case   $\tau$ is no
longer purely imaginary, but should retain  a positive imaginary
part)

By  integrating $d/dz_i [\ln f(z_1, \ldots, z_{N_e})]$ around the
boundaries of the rectangle, and using  Eq.~(\ref{EQ:BC2}) to
combine the contributions of the opposite sides, we see  that the
number of zeros of $f$, considered as a function of $z_i$,  is
$N_s$. The precise locations of these $N_s$ zeros will depend on the
positions of the the other $z_j$, but  they are subject to a
non-trvial constraint \cite{stone_book}.  Suppose that $g(z)$ is
meromophic and doubly-periodic: \be g(z)=g(z+1)=g(z+\tau).
 \ee
 By evaluating  the integral
\begin{equation}
I = \frac{1}{2\pi i} \oint z \frac{g'(z)}{g(z)} dz \label{EQ:Abel1}
\end{equation} around the edges of the
period parallelogram $0 \to 1 \to 1+\tau \to \tau\ \to 0$, and again
combining the contributions from opposite sides, we find that \be I=
m+n\tau \label{EQ:Abel2} \ee where $m$ and $n$ are integers. On the
other hand, if the poles of $g(z)$ in the period parallelogram are
at $z = b_i$ and the zeros at $z = a_i$, then $I = \sum_i a_i -
\sum_i b_i$. Taken together with Eq.~(\ref{EQ:Abel2}), this  means
that the sum of zeros minus the sum of poles  vanishes modulo
periods. This is {\it Abel's theorem\/} for the torus \cite{tata}.
Although $f(z_1, \ldots, z_{N_e})$ is {it not\/}   doubly periodic,
the ratio \be g(z)\equiv f(z, ,z_2\ldots, z_{N_e})/f(z,z_2' \ldots,
z'_{N_e}) \ee
 is a doubly-periodic meromorphic function whose zeros are at the zeros
of $f(z,z_2\ldots, z_{N_e})$  and whose poles are at the zeros  of
$f(z,z_2' \ldots, z'_{N_e})$ . Consequently the sum of $N_s$ zeros
of $f(z_1,z_2,\ldots, z_{N_e})$ considered as a function of  $z_1$
is independent of other electron coordinates. The same is true of
the zeros of $f$ considered as a function of any of the $z_i$.

The defining characteristic of a Laughlin wave function at filling
fraction $1/q$ (where $q$ is an odd integer and  $N_s = qN_e$) is
that it vanishes as $(z_i-z_j)^q$ as any $z_i$ approaches any other
$z_j$. When combined with the condition discussed in the last
paragraph, namely that the sum of all zeros for each $z_i$ should be
constant, the only possible form of the holomorphic part of the wave
function for the Laughlin state on the torus is
\cite{HR_deg_wf,stone_book}
\begin{equation}
f(z_1, z_2, \ldots, z_{N_e}) = F_{{\rm cm}} \Big(\sum_i z_i\Big)
\prod_{i<j} [\vartheta_1 (z_i - z_j)]^q. \label{EQ:Laughlin1}
\end{equation}
Here $F_{{\rm cm}} (Z)$ is a holomorphic function possessing  $q$
zeros, and $\vartheta_1(z)$ is one of the four  Jacobi theta
functions:
\begin{eqnarray}
\vartheta_1 (z) &=& - \vartheta \left [
\begin{array}{c}
  1/2 \\
  1/2 \\
\end{array}
\right ] (z\,|\,\tau),\nonumber\\
\vartheta_2 (z) &=&\phantom - \vartheta \left [
\begin{array}{c}
  1/2 \\
  0 \\
\end{array}
\right ] (z\,|\,\tau),\nonumber\\
\vartheta_3 (z) &=& \phantom -\vartheta \left [
\begin{array}{c}
\,\,  \,0\,\,\, \\
 \,\, \,0\,\,\, \\
\end{array}
\right ] (z\,|\,\tau),\nonumber\\
\vartheta_4 (z) &=&\phantom - \vartheta \left [
\begin{array}{c}
  0 \\
  1/2 \\
\end{array}
\right ] (z\,|\,\tau), \label{EQ:JThetas}
\end{eqnarray}
where the {\it theta function with characteristics\/} is defined as
\begin{equation}
\vartheta \left [
\begin{array}{c}
  a \\
  b \\
\end{array}
\right ] (z\,|\,\tau) = \sum_{n=-\infty}^{\infty} \exp \, [i \pi
\tau (n+a)^2 + 2 \pi i (n+a)(z+b)]. \label{EQ:theta0}
\end{equation}
We will usually write the theta function with characteristics as
$\vartheta[\alpha](z,\tau)$ where $\alpha$ is the vector $(a,b)^T$.

Each of the theta  functions possesses  a single zero in the period
parallelogram, but note that for $a,b \in {\mathbb Z}/2$ \be
\vartheta[\alpha](-z,\tau)= (-1)^{4ab}\vartheta[\alpha](z,\tau), \ee
so only $\vartheta_1$ obeys $\vartheta_1(-z)=-\vartheta_1(z)$ and
has its zero at $z=0$. All theta functions are quasi-periodic, and,
in particular,
\begin{eqnarray}
\vartheta_1 (z+1) &=& -\vartheta_1 (z), \nonumber\\
\vartheta_1 (z+\tau) &=& -\exp[-i \pi (2z + \tau)] \vartheta_1 (z).
\label{EQ:theta1BC}
\end{eqnarray}
Eqs.~(\ref{EQ:BC2}), (\ref{EQ:Laughlin1}), and (\ref{EQ:theta1BC}),
together with $N_s = qN$, require the following quasi-periodicity
conditions for $F_{{\rm cm}} (Z)$: \cite{HR_deg_wf}
\begin{eqnarray}
F_{{\rm cm}} (Z+1) &=& (-1)^{(N_s - q)} F_{{\rm cm}} (Z), \nonumber\\
F_{{\rm cm}} (Z+\tau) &=& (-1)^{(N_s - q)} \exp[-i \pi q (2z +
\tau)] F_{{\rm cm}} (Z). \label{EQ:cmBC}
\end{eqnarray}
A convenient  basis for such $F_{{\rm cm}} (Z)$  is  provided by the
set of $q$ functions \cite{read-rezayi_paired, kol-read}
\begin{equation}
F_{{\rm cm}}^{(m)} (Z) = \vartheta \left [
\begin{array}{c}
  m/q + (N_s - q)/2q \\
  - (N_s - q)/2 \\
\end{array}
\right ] (qZ\,|\,q\tau), \label{EQ:cmWF}
\end{equation}
where $m$ is an integer defined mod $q$. The  resulting set of $q$
linearly-independent   wave functions differ only by a rigid
translation  along the $y$ direction, and in the limit
 $N_e \rightarrow \infty$ each  electron need  move only
infinitesimal amount to cause this shift. The distinct wave functions
are locally indistinguishable, and  local perturbations change the 
energy of these states by same amount. The $q$-fold  degeneracy is 
therefore unaffected by such local perturbations.

If  now $N_h$ quasiholes are inserted at $w_i$, then $N_s = qN_e +
N_h$. Since $f$ should now vanish to the first power as $z_i
\rightarrow w_j$, the only way to have the sum of the zeros of each
$z_i$ independent of $\{w_j\}$ is to set \cite{HR_deg_wf,stone_book}
\begin{equation}
f^{(m)}(z_1, z_2, \ldots, z_{N_e}) = F^{(m)}_{{\rm cm}} \left(
\sum_i z_i + \frac{\sum_j w_j}{q} \right) \prod_{i,j}
\vartheta_1(z_i - w_j) \prod_{i<j} [\vartheta_1 (z_i - z_j)]^q.
\label{EQ:LaughlinQH1}
\end{equation}
It is useful to include a purely  $\{w_j\}$-dependent part to  the
wave function normalization so that
\begin{eqnarray}
\Psi^{(m)}({\bf r}_1, \ldots, {\bf r}_{N_e}; {\bf R}_1, \ldots, {\bf
R}_{N_h} ) &=& \exp  \left(-\sum_i \frac{y_i^2}{2 l^2}\right) \exp
\left(-\sum_j \frac{\eta_j^2}{2ql^2}\right) F^{(m)}_{{\rm cm}}
\left( \sum_i z_i + \frac{\sum_j w_j}{q}
\right)\nonumber\\
&\times& \prod_{i<j} [\vartheta_1 (w_i - w_j)]^{1/q} \prod_{i,j}
\vartheta_1(z_i - w_j) \prod_{i<j} [\vartheta_1 (z_i - z_j)]^q,
\label{EQ:LaughlinQH2}
\end{eqnarray}
where $w_i = (\xi_i + i\eta_i)/L_x$ and $\textbf{R} = (\xi, \eta)$.
It was argued by Einarsson \cite{einarsson} that taking one
quasihole around another on torus should result in the same phase
factor on a torus as on plane and this factor is accounted  for by
$\prod_{i<j} [\vartheta_1 (w_i - w_j)]^{1/q}$. The exponent of the
new gaussian factor can be explained from the fact that the charge
of a quasihole is $e/q$. These added factors in
Eq.~(\ref{EQ:LaughlinQH2}) should therefore have converted   all
Berry phases into explicit monodromies.

One can easily check that for $\{z_i\}$, boundary conditions of
Eq.~(\ref{EQ:LaughlinQH2}) is exactly that of Eq.~(\ref{EQ:BC1}),
provided that $N_s$ of Eq.~(\ref{EQ:cmBC}) is  equal to $qN_e +
N_h$.  Under a straight-line analytic continuation $z\to z\pm 1$ we
have \be \left[\vartheta_1(z )\right]^{1/q} \to  \exp (\pm i\pi/q)
[\vartheta_1(z)]^{1/q} \ee and as $z\to z\pm \tau$ we have \be
\left[\vartheta_1(z)\right] ^{1/q} \to  \exp(\mp i\pi/q) \exp \left[
-\frac{i \pi}{q}(\pm 2z + \tau) \right] [\vartheta_1(z)]^{1/q}. \ee
Also note for the center-of-mass wave function:
\begin{eqnarray}
F_{{\rm cm}}^{(m)} (Z \pm 1/q) & = & (-1)^{N_e} e^{\pm \pi i (N_h -
q)/q} \exp\left[\pm 2\pi i \frac{m}{q}\right]F_{{\rm cm}}^{(m)}
(Z),\nonumber\\
F_{{\rm cm}}^{(m)} (Z \pm \tau/q) & = & (-1)^{N_e} e^{\pm \pi i (N_h
- q)/q} \exp[-\pi i (\pm 2z + \tau/q)]F_{{\rm cm}}^{(m \pm 1)} (Z).
\label{EQ:cmTrans}
\end{eqnarray}

Taking a quasihole around one of the generators of the torus
therefore effects the following transformations: under
$\textbf{R}_i\to \textbf{R}_i \pm L_x {\bf {\hat x}}$ we have \bea
\Psi^{(m)}({\bf r}_1, \ldots, {\bf r}_{N_e}; {\bf R}_1, \ldots, {\bf
R}_i , \ldots ) &\to&  \exp \,\left[ \pm 2\pi i \left(\frac{m}{q} +
\frac{2N_h-q-1}{2q}\right) \right]\nonumber\\ &\times&
\Psi^{(m)}({\bf r}_1, \ldots, {\bf r}_{N_e}; {\bf R}_1, \ldots, {\bf
R}_i, \ldots ), \label{EQ:transQH1} \eea and under $  {\bf R}_i \to
{\bf R}_i \pm L_y{\bf {\hat y}}$ we have \bea \Psi^{(m)}({\bf r}_1,
\ldots, {\bf r}_{N_e}; {\bf R}_1, \ldots, {\bf R}_i , \ldots ) &\to&
-e^{\pm i \pi/q} \exp \, (\mp i L_y \xi_i / ql^2)\nonumber\\
&\times& \Psi^{(m \pm 1)}({\bf r}_1, \ldots, {\bf r}_{N_e}; {\bf
R}_1, \ldots, {\bf R}_i, \ldots ). \label{EQ:transQH2} \eea
 Eqs.~(\ref{EQ:transQH1}),
(\ref{EQ:transQH2})  reproduce the operator transformation of Wen
and Niu \cite{wen-niu} up to  a gauge transformation: motion of the
quasihole about one torus generator reproduces the state up to
phase, and motion about the other generator rolls the ground state
over into another one \footnote{The constant factor $-\exp(\pm i
\pi/q)$ can be taken care of if the ground state wave function is
re-defined with suitable phase factor.}. These operations and their
effects are  analogous  to the operations appearing in the Verlinde
algebra of conformal field theory.

\section{Moore-Read state}

The holomorphic part of the $\nu=1/2$ Moore-Read state wave function
on the plane is \be f_{\rm{MR}}(z_1,\ldots, z_{2n})= {\rm
Pf}\left(\frac{1}{z_i-z_j}\right)\prod_{i<j}(z_i-z_j)^2,
 \ee where
the Pfaffian of a $2n$-by-$2n$ antisymmetric matrix is defined as
 \be
 {\rm Pf}(A)= \frac{1}{2^nn!} \sum_{P \in S_{2n}} {\rm sgn} (P)
\prod_{i=1}^{n} A_{P(2i-1),P(2i)}.
 \ee
Here $P$ runs over all permutation of $2n$ objects. This wave
function is an antisymmetric polynomial in the $z_i$, the poles in
the Pfaffian part having canceled against some of the zeros in the
Laughlin-Jastrow factor. The suppression of these zeros by  the
Pfaffian factor  increases the amplitude  for the one electron to
approach another. The Pfaffian  can therefore be considered to
indicate a pairing between electrons \cite{greiter-wen}, and this
pairing  is closely related to the  BCS pairing in $p+ip$
superconductors \cite{read-green}.


In constructing the  Moore-Read wave functions on a torus we must
satisfy the boundary conditions Eq.~(\ref{EQ:BC2}).  These boundary
conditions come from the   gauge choice and the fact that all
electrons are in the lowest Landau level;  they do not depend on
what correlated many-body state the electrons are in.

Now the  Pfaffian part is  equal to the correlator of $2n$ chiral
Majorana fermion fields of the critical Ising model \be \langle
\psi(z_1) \ldots \psi(z_{2n})\rangle = {\rm
Pf}\left(\frac{1}{z_i-z_j}\right).\label{EQ:Psipair} \ee There is
more than one way to put such a correlator on a torus. This is
because we are free to give the Ising fields periodic or
antiperiodic boundary conditions around each generator. We will see
that there is an intricate  interplay between the boundary
conditions  (\ref{EQ:BC2}) required of  the physical electrons and
the boundary condition choices we make for the Ising  $\psi$'s.

\subsection{Even-spin structure sector}

The Laughlin-Jastrow  part of the wave function on the torus can be
dealt with by the same $(z_i-z_j)\to \vartheta_1(z_i-z_j)$
substitution as before.  The $1/(z_i-z_j)$ factors in the Pfaffian
require a more careful treatment. Because the Pfaffian contains {\it
sums\/}  of products any  $z$-dependent factors arising from the  $z
\rightarrow z+1$ or $z \rightarrow z+\tau$ quasi-periodicity
properties  will not appear as an overall common factor.
Consequently we should arrange that there are {\it no} $z$-dependent
factors at all, and simply setting  $1/(z_i - z_j) \to
1/\vartheta_1(z_i - z_j)$ will not do. A possible choice is to set
\cite{greiter-wen}
\begin{equation}
\frac{1}{z_i-z_j} \to  \frac{\vartheta[\alpha](z_i -
z_j)}{\vartheta_1(z_i - z_j)} \label{EQ:Pfaffian}
\end{equation}
for some choice of $\alpha=(1/2,0)^T$, $(0,0)^T$, $(0,1/2)^T$. The
functions $\vartheta[\alpha](z)/\vartheta_1(z)$ now obey the
following boundary conditions:
\begin{eqnarray}
\frac{\vartheta[\alpha](z+1)}{\vartheta_1(z+1)} &=& -e^{2\pi i
a} \frac{\vartheta[\alpha](z)}{\vartheta_1(z)},\nonumber\\
\frac{\vartheta[\alpha](z+\tau)}{\vartheta_1(z+\tau)}&=& -e^{2\pi i
b} \frac{\vartheta[\alpha](z)}{\vartheta_1(z)}. \label{EQ:PfaffBC}
\end{eqnarray}
These functions are proportional to the two-point functions \be
\vev{\psi(z)\psi(z')}_{\alpha}=
\frac{\vartheta'_1(0)\vartheta[\alpha](z- z')}{\vartheta[\alpha](0)
\vartheta_1(z-z')}, \label{EQ:MajoranaBC} \ee for chiral Majorana
fermions with different spin  structures --- {\it i.e.\/} with
different periodic or antiperiodic boundary conditions round the
generators of the torus. At least one of these boundary conditions
must be antiperiodic. If this condition is not met,  the fermion
propagator has a zero mode, and the usual two-point function does
not exist. We are therefore, at the moment considering only {\it
even\/} spin structures, {\it i.e.\/} spin structures
$\alpha=(a,b)^T$ having $4ab$ an even integer. It should be noted
once again that these periodic and antiperiodic boundary conditions
are merely properties of ingredients in the Pfaffian. We are {\it
not\/} changing the boundary conditions of the many-body wave
function. Instead the sign factors are  accommodated by the
center-of-mass wave function. This means that the center-of-mass
wave functions obtained by setting $q=2$ in Eq.~(\ref{EQ:cmWF}),
\begin{eqnarray}
{\tilde F}_{{\rm cm}}^{(m)} (Z) &=& \vartheta \left [
\begin{array}{c}
  m/2 + (N_s - 2)/4 \\
  - (N_s - 2)/2\\
\end{array}
\right ] (2Z\,|\,2\tau), \label{EQ:cmWFPfaffOdd}
\end{eqnarray}
will  no longer suffice. Instead there are  distinct  center-of-mass
wave functions for each  $\alpha = (a,b)^T$:
\begin{equation}
F_{{\rm cm}}^{(\alpha,m)} (Z) = \vartheta \left [
\begin{array}{c}
  m/2 + (N_s - 2)/4 + (1-2a)/4\\
  - (N_s - 2)/2 - (1-2b)/2\\
\end{array}
\right ] (2Z\,|\,2\tau). \label{EQ:cmWFPfaff}
\end{equation}
Here $m$ is an integer defined mod 2. (It will be shown later that
Eq.~(\ref{EQ:cmWFPfaffOdd}) {\it is\/} the center-of-mass wave
function in the {\it odd} spin structure.) The complete wave
function
\begin{equation}
\Psi^{(\alpha,m)}({\bf r}_1, \ldots, {\bf r}_{N_e}) = \exp \,
(-\sum_i y_i^2/2 l^2) F^{(\alpha,m)}_{{\rm cm}} (\sum_i z_i) {\rm
Pf} \left(\frac{\vartheta[\alpha](z_i - z_j)}{\vartheta_1(z_i -
z_j)}\right) \prod_{i<j} [\vartheta_1 (z_i - z_j)]^2,
\label{EQ:PfaffWF}
\end{equation}
with $\alpha=(1/2,0)^T$, $(0,0)^T$, or $(0,1/2)^T$, now satisfies
the boundary conditions Eq.~(\ref{EQ:BC2}). Note that the additional
$\alpha$-dependent terms that Eq.~(\ref{EQ:cmWFPfaff}) posesses  in
comparison with Eq.~(\ref{EQ:cmWFPfaffOdd}) ensures that that the
boundary conditions of the total wave functions are same for
different $\alpha$'s. The three spin-structure choices coupled with
the two possible values of $m$ in Eq.~(\ref{EQ:PfaffWF}) give us the
sixfold degeneracy of the even-spin structure  Moore-Read state on
the torus \cite{masaki,greiter-wen,read-green}.

Now we consider inserting quasiholes.   A charge $e/2$ quasihole
excitation in one of these degenerate ground state is little
different from a quasihole in a Laughlin state. Such a quasihole has
one quantum of flux and consequently $N_s = 2N_e +1 $. When  an
$e/2$  quasihole is inserted  at $w$, the holomorphic part of the
wave function  becomes
\begin{equation}
f^{(\alpha,m)}(z_1, \ldots, z_{N_e}; w) = F^{(\alpha,m)}_{{\rm cm}}
\left( \sum_i z_i + \frac{w}{2} \right) {\rm Pf}
\left(\frac{\vartheta[\alpha](z_i - z_j)}{\vartheta_1(z_i -
z_j)}\right) \prod_i \vartheta_1 (z_i - w) \prod_{i<j} [\vartheta_1
(z_i - z_j)]^2.
\label{EQ:fullQH}
\end{equation}
When this quasihole is carried around the generators, the wave
function Eq.~(\ref{EQ:fullQH}) behaves almost  the same as the torus
Laughlin wave function. Under $w \rightarrow w \pm 1$ we have \be
f^{(\alpha,m)}(z_1, \ldots, z_{N_e}; w) \to
\begin{cases}
(-1)^m f^{(\alpha,m)}(z_1, \ldots, z_{N_e}; w), & \alpha = \left(
\begin{array}{c}
  0 \\
  0 \\
\end{array}
\right), \left (
\begin{array}{c}
  0 \\
  \frac{1}{2} \\
\end{array}
\right )\\
\mp i (-1)^m f^{(\alpha,m)}(z_1, \ldots, z_{N_e}; w), & \alpha =
\left (
\begin{array}{c}
  \frac{1}{2} \\
  0 \\
\end{array}
\right )
\end{cases} \ee
and under $w \rightarrow w \pm \tau$ we have \be
f^{(\alpha,m)}(z_1,\ldots, z_{N_e}; w) \to
\begin{cases}
\exp \left[- \frac{i \pi N_s}{2} (\pm 2 w + \tau) \right]
f^{(\alpha,m+1)}(z_1, \ldots, z_{N_e};w), & \alpha = \left (
\begin{array}{c}
  \frac{1}{2} \\
  0 \\
\end{array}
\right ), \left(
\begin{array}{c}
  0 \\
  0 \\
\end{array}
\right)\\
\mp i \exp \left[- \frac{i \pi N_s}{2} (\pm 2 w + \tau) \right]
f^{(\alpha,m+1)}(z_1, \ldots, z_{N_e}; w), & \alpha = \left (
\begin{array}{c}
  0 \\
  \frac{1}{2} \\
\end{array}
\right ).
\end{cases}
\label{EQ:FullPfaffBC} \ee This is  to be expected because creating
the $e/2$  quasihole breaks no pairs and so has  no effect on the
BCS pairing characterizing  the Moore-Read state. (These
transformations have been discussed by Oshikawa {\it et al.} in the
operator language \cite{masaki}. There, the monodromy  was
associated with the adiabatic insertion of a  unit flux quantum into
the ``holes'' of the torus.)

The $e/2$ quasihole is not, however,  the elementary excitation for
the Moore-Read state. By allowing pair-breaking,  a charge $e/2$
quasihole with one quantum flux can  fractionalize into two
charge-$e/4$ quasiholes, each possessing   a {\it half\/} quantum of
flux \cite{moore-read,read-green}.  To construct  a wave function
with two charge $e/4$ quasiholes, the Pfaffian part needs to be
modified, as this is the part that describes  the BCS pairing.

There are three basic conditions to be considered when writing down
such a two-quasihole wave function. The first is that when these two
quasiholes are brought together we should recover
Eq.~(\ref{EQ:fullQH}). The second condition is that if  two {\it
paired} electrons wind around a single  charge $e/4$ quasihole, it
should result in accumulation of phase $2\pi$. The third condition
is that, since the argument of the center-of-mass wave function
becomes  $\sum_i z_i + (w_1 + w_2)/4$, the boundary condition of
the Pfaffian part for the  $z_i \rightarrow z_i + \tau$ translation
needs to change accordingly. Greiter {\it et al.} obtained
wave functions satisfying these three conditions \cite{greiter-wen}.
The holomorphic part of their wave function is
\begin{equation}
f^{(\alpha,m)}(z_1, \ldots, z_{N_e}; w_1, w_2) = F^{({\bf
\alpha},m)}_{{\rm cm}} \left( \sum_i z_i + \frac{w_1 + w_2}{4}
\right) {\rm Pf} \left(M^{\alpha}_{ij}\right) \prod_{i<j}
[\vartheta_1 (z_i - z_j)]^2, \label{EQ:halfQH}
\end{equation}
with
\begin{equation}
M^{\alpha}_{ij} = \frac{\vartheta[\alpha](z_i - z_j + w_{12}/2)
\vartheta_1 (z_i - w_1) \vartheta_1 (z_j - w_2) + (i \leftrightarrow
j)}{2\vartheta_1(z_i - z_j)}, \label{EQ:frac}
\end{equation}
and  $w_{12} = w_1 - w_2$. (Note that $(i \leftrightarrow j)$ refers
to a term that differs from the previous term only by the exchange of the index
$i$ and $j$; hence, for Eq.~(\ref{EQ:frac}), $(i \leftrightarrow j) \equiv
\vartheta[\alpha](z_j - z_i + w_{12}/2) \vartheta_1 (z_j - w_1) \vartheta_1 (z_i - w_2)$.)
As far as translation of electrons around the torus generators are
concerned Eq.~(\ref{EQ:halfQH}) has same transformation as
Eq.~(\ref{EQ:fullQH}), or as the holomorphic part of Eq.~(\ref{EQ:PfaffWF}).

The above considerations are {\it not}, however, sufficient for
obtaining the complete two-quasihole wave functions.
 Eq.~(\ref{EQ:halfQH}) and Eq.~(\ref{EQ:frac}) satisfy  constraints
involving only the electron co-ordinates $z_i$. There are also
important  factors that depend only on $w_1$ and $w_2$. To obtain
the complete  dependence on the quasihole coordinates as well, we
need to calculate the holomorphic conformal blocks of the
correlators for critical Ising model.

The starting point is to observe from Moore and Read's original
derivation how the Pfaffian part of the Moore-Read state wave
function with two quasiholes is obtained when place on the plane
\cite{moore-read}:
\begin{eqnarray}
{\rm Pf} \left(\frac{(z_i - w_1)(z_j - w_2) + (i \leftrightarrow
j)}{z_i - z_j}\right) &=& \langle \psi(z_1) \cdots \psi(z_{N_e})
\sigma(w_1) \sigma(w_2) \rangle \nonumber\\
&\times& w_{12}^{1/8} \prod_i [(z_i - w_1)(z_i - w_2)]^{1/2},
\end{eqnarray}
where $\psi$ is the Majorana fermion and $\sigma$ the spin field.
One can immediately see that the Pfaffian part of the wave function
on a torus would be equal to
\begin{equation}
\langle \psi(z_1) \cdots \psi(z_{N_e}) \sigma(w_1) \sigma(w_2)
\rangle_{\alpha} [\vartheta_1(w_{12})]^{1/8} \prod_i
[\vartheta_1(z_i - w_1) \vartheta_1(z_i - w_2)]^{1/2}.
\label{EQ:PfaffCorrel}
\end{equation}
Here $\alpha = (a,b)^T$ denotes the boundary conditions,
Eq.~(\ref{EQ:PfaffBC}), of the $\psi$ field on the torus, just as in
the case of Eq.~(\ref{EQ:MajoranaBC}).

The following Ising model correlators have been obtained in the even
spin structures on the torus by Di Francesco {\it et al.}
\cite{francesco,francesco_paper}
\begin{eqnarray}
\frac{\langle \psi(z_i) \psi(z_j) \sigma(w_1)
\sigma(w_2)\rangle_{\alpha}}{\langle \sigma(w_1)
\sigma(w_2)\rangle_{\alpha}} &=& \frac{\vartheta'_1(0)}{2
\vartheta_1
(z_i - z_j)}\nonumber\\
&\times& \left[\frac{\vartheta[\alpha](z_i - z_j +
w_{12}/2)}{\vartheta[\alpha](w_{12}/2)}\left(\frac{\vartheta_1(z_i -
w_1)\vartheta_1(z_j - w_2)}{\vartheta_1(z_i - w_2)\vartheta_1(z_j -
w_1)}\right)^{1/2} + (i \leftrightarrow j)\right],\nonumber\\
\langle \sigma(w_1) \sigma(w_2)\rangle_{\alpha} &=&
\left(\frac{\vartheta[\alpha](w_{12}/
2)}{\vartheta[\alpha](0)}\right)^{1/2}
\left(\frac{\vartheta'_1(0)}{\vartheta_1 (w_{12})}\right)^{1/8}.
\label{EQ:IsingCorrel}
\end{eqnarray}
(To be precise, these are the chiral holomorphic parts extracted
from the non-chiral Ising field correlators obtained by Di Francesco
{\it et al.}.)  From the antisymmetry of the $\psi$ field under
exchange, and from the condtitions
\begin{eqnarray}
\lim_{z_1 \rightarrow z_2} (z_1 - z_2) \langle \psi(z_1) \psi(z_2)
\psi(z_3) \cdots \psi(z_{N_e}) \sigma(w_1) \sigma(w_2)
\rangle_{\alpha} &=& \langle \psi(z_3) \cdots \psi(z_{N_e})
\sigma(w_1) \sigma(w_2) \rangle_{\alpha},\nonumber\\
\lim_{z_i \rightarrow z_j} (z_i - z_j) \frac{\langle \psi(z_i)
\psi(z_j) \sigma(w_1) \sigma(w_2)\rangle_{\alpha}}{\langle
\sigma(w_1) \sigma(w_2)\rangle_{\alpha}} &=& 1, \label{EQ:IsingLim}
\end{eqnarray}
we  can obtain  the $N_e$-point  $\psi$ field correlator ($N_e$
even)  as
\begin{equation}
\langle \psi(z_1) \cdots \psi(z_{N_e}) \sigma(w_1) \sigma(w_2)
\rangle_{\alpha} = \langle \sigma(w_1) \sigma(w_2)\rangle_{\alpha}
{\rm Pf} \left(\frac{\langle \psi(z_i) \psi(z_j) \sigma(w_1)
\sigma(w_2)\rangle_{\alpha}}{\langle \sigma(w_1)
\sigma(w_2)\rangle_{\alpha}}\right). \label{EQ:manyIsingCorrel}
\end{equation}
Eqs.~(\ref{EQ:PfaffCorrel}), (\ref{EQ:IsingCorrel}), and
(\ref{EQ:manyIsingCorrel}) tell us that the wave functions of
Eq.~(\ref{EQ:halfQH}) need to be modified in the following way if
they are  to have correct dependence on quasihole coordinates:
\begin{eqnarray}
f^{(\alpha,m)}(z_1, \ldots, z_{N_e}; w_1, w_2) &=&
F^{(\alpha,m)}_{{\rm cm}} \left( \sum_i z_i + \frac{w_1 + w_2}{4}
\right) [\vartheta[\alpha](w_{12}/2)]^{1/2} {\rm Pf} \left({\tilde
M}^{\alpha}_{ij}\right) \nonumber\\
&\times&  \prod_{i<j} [\vartheta_1 (z_i - z_j)]^2,
\label{EQ:halfQH1}
\end{eqnarray}
with
\begin{equation}
{\tilde M}^{\alpha}_{ij} = \frac{\vartheta[\alpha](z_i - z_j +
w_{12}/2) \vartheta_1 (z_i - w_1) \vartheta_1 (z_j - w_2) + (i
\leftrightarrow j)}{2\vartheta_1(z_i - z_j)
\vartheta[\alpha](w_{12}/2)}. \label{EQ:frac1}
\end{equation}
Note that while Eq.~(\ref{EQ:IsingCorrel}) is {\it not}
single-valued in coordinates of the $\psi$ fields---taking a $\psi$
around a $\sigma$ results in a minus sign---the wavefunctiuon
Eq.~(\ref{EQ:halfQH1}) {\it is} single-valued in the electron
coordinate. This is  because the $[\vartheta_1
(w_{12})]^{1/8}\prod_i [\vartheta_1(z_i - w_1) \vartheta_1(z_i -
w_2)]^{1/2}$ factor (in addition to the Laughlin-Jastrow factor)
makes everything analytic in Eq.~(\ref{EQ:halfQH1}), save for
$[\vartheta[\alpha](w_{12}/2)]^{1/2}$. Thus the wave function is
single-valued in electron coordinate, and has no singularity. Note
that constants  $\vartheta[\alpha](0)$ and $\vartheta'_1(0)$ have
been  ignored in Eq.~(\ref{EQ:halfQH1}). (This means in the limit
$w_1 \to w_2$ Eq.~(\ref{EQ:halfQH1}) will differ from
Eq.~(\ref{EQ:fullQH}) by some multiplicative constant.)

We now ask what  happens when  one of the quasiholes is translated
around the generators. In contrast to Eq.~(\ref{EQ:FullPfaffBC})
where only $m$ changed, this operation  will  result in a change in
$\alpha$. 
To see what happens to the  wave functions, we require  the following
standard theta function identities \cite{francesco,tata}:
\begin{eqnarray}
\vartheta_2 (z \pm 1/2) &=& \mp \vartheta_1 (z), \nonumber\\
\vartheta_2 (z \pm \tau/2) &=& \exp [-i \pi (\pm z + \tau/4)]
\vartheta_3 (z), \nonumber\\
\vartheta_3 (z \pm 1/2) &=& \vartheta_4 (z), \nonumber\\
\vartheta_3 (z \pm \tau/2) &=& \exp [-i \pi (\pm z + \tau/4)]
\vartheta_2 (z), \nonumber\\
\vartheta_4 (z \pm 1/2) &=& \vartheta_3 (z), \nonumber\\
\vartheta_4 (z \pm \tau/2) &=& \pm i \exp [-i \pi (\pm z + \tau/4)]
\vartheta_1 (z), \label{EQ:thetaTrans}
\end{eqnarray}
together with the center-of-mass wave function formulas,
\begin{eqnarray}
F_{{\rm cm}}^{(a=2,m)} (z \pm 1/4) &=& e^{-i(1/2 \mp 1/2)\pi(m +
N_s/2 -1)} {\tilde F}_{{\rm cm}}^{(m)} (z), \nonumber\\
F_{{\rm cm}}^{(a=2,m)} (z \pm \tau/4) &=& e^{\pm i\pi(N_s - 1)/4}
\exp [-i \pi(\pm z + \tau/8)] F_{{\rm cm}}^{(a=3,m+ 1/2 \mp 1/2)}
(z),\nonumber\\
F_{{\rm cm}}^{(a=3,m)} (z \pm 1/4) &=& e^{-i(1/2 \mp 1/2)\pi(m +
N_s/2 - 1/2)} F_{{\rm cm}}^{(a=4,m)} (z), \nonumber\\
F_{{\rm cm}}^{(a=3,m)} (z \pm \tau/4) &=& e^{\pm i\pi(N_s - 1)/4}
\exp [-i \pi(\pm z + \tau/8)] F_{{\rm cm}}^{(a=2,m+ 1/2 \pm 1/2)}
(z),\nonumber\\
F_{{\rm cm}}^{(a=4,m)} (z \pm 1/4) &=& e^{i(1/2 \pm 1/2)\pi(m +
N_s/2 - 1/2)} F_{{\rm cm}}^{(a=3,m)} (z), \nonumber\\
F_{{\rm cm}}^{(a=4,m)} (z \pm \tau/4) &=& e^{\pm i\pi(N_s - 2)/4}
\exp [-i \pi(\pm z + \tau/8)] {\tilde F}_{{\rm cm}}^{(m + 1/2 \pm
1/2)} (z). \label{EQ:cmTrans}
\end{eqnarray}
In these center-of-mass wave function formulas, the notation of
Eq.~(\ref{EQ:JThetas}) is used; that is, $a = 2$ stands for $\alpha
= (1/2,0)^T$, $a = 3$ for $\alpha = (0,0)^T$, and $a = 4$ for
$\alpha = (0,1/2)^T$.

Note that the transformations of Eq.~(\ref{EQ:thetaTrans}) are
covariant with those of Eq.~(\ref{EQ:cmTrans}).

Consider the results not involving $\vartheta_1$ or ${\tilde
F}_{{\rm cm}}^{(m)}$:
\begin{eqnarray}
w_1 \to w_1 \pm \tau : f^{(a=2,m)}(z_1, \ldots, z_{N_e}; w_1, w_2)
&\to& \exp [-i
\pi N_s (\pm w_1/2 + \tau/4)] \nonumber\\
&\times& f^{(a=3,m + 1/2 \mp 1/2)}(z_1, \ldots, z_{N_e}; w_1,
w_2),\nonumber\\
w_1 \to w_1 \pm 1 : f^{(a=3,m)}(z_1, \ldots, z_{N_e}; w_1, w_2)
&\to& (-1)^{N_e/2}
e^{i \pi m (1/2 \mp 1/2)} \nonumber\\
&\times& f^{(a=4,m)}(z_1, \ldots, z_{N_e}; w_1, w_2),\nonumber\\
w_1 \to w_1 \pm \tau : f^{(a=3,m)}(z_1, \ldots, z_{N_e}; w_1, w_2)
&\to& \exp [-i
\pi N_s (\pm w_1/2 + \tau/4)] \nonumber\\
&\times& f^{(a=2,m + 1/2 \pm 1/2)}(z_1, \ldots, z_{N_e}; w_1,
w_2),\nonumber\\
w_1 \to w_1 \pm 1 : f^{(a=4,m)}(z_1, \ldots, z_{N_e}; w_1, w_2)
&\to& (-1)^{N_e/2}
e^{i \pi m (1/2 \pm 1/2)} \nonumber\\
&\times& f^{(a=3,m)}(z_1, \ldots, z_{N_e}; w_1, w_2).
\label{EQ:halfTrans}
\end{eqnarray}
In  Moore and Read's original wave function formulation
\cite{moore-read} the Gaussian factor comes entirely determined from
the charge sector. Since the charge of a quasihole considered here
is $e/4$, an  analogy with Eq.~(\ref{EQ:LaughlinQH2}) shows that the
total wave function should be
\begin{equation}
\Psi^{(\alpha,m)}({\bf r}_1, \ldots, {\bf r}_{N_e}; {\bf R}_1, {\bf
R}_2 ) = \exp \, (-\sum_i y_i^2/2 l^2) \exp \, [- (\eta_1^2 +
\eta_2^2)/8l^2] \, f^{(\alpha,m)}(z_1, \ldots, z_{N_e}; w_1, w_2).
\label{EQ:MR2qh}
\end{equation}
Eq.~(\ref{EQ:halfTrans}) and Eq.~(\ref{EQ:MR2qh}) gives the
following transformation rules for the total wave functions:
\begin{eqnarray}
{\bf R}_1 \to {\bf R}_1 \pm L_y {\bf {\hat y}} : \Psi^{(a=2,m)}({\bf
r}_1, \ldots, {\bf r}_{N_e}; {\bf R}_1, {\bf R}_2 ) &\to& \exp \,
(\mp i L_y
\xi_1 / 4l^2) \nonumber\\
&\times& \Psi^{(a=3, m +1/2 \mp 1/2)}({\bf r}_1, \ldots, {\bf
r}_{N_e}; {\bf R}_1, {\bf R}_2), \nonumber\\
{\bf R}_1 \to {\bf R}_1 \pm L_x {\bf {\hat x}} : \Psi^{(a=3,m)}({\bf
r}_1, \ldots, {\bf r}_{N_e}; {\bf R}_1, {\bf R}_2 ) &\to&
(-1)^{N_e/2} e^{i \pi
m (1/2 \mp 1/2)} \nonumber\\
&\times& \Psi^{(a=4,m)}({\bf r}_1, \ldots, {\bf r}_{N_e};
{\bf R}_1, {\bf R}_2 ), \nonumber\\
{\bf R}_1 \to {\bf R}_1 \pm L_y {\bf {\hat y}} : \Psi^{(a=3,m)}({\bf
r}_1, \ldots, {\bf r}_{N_e}; {\bf R}_1, {\bf R}_2 ) &\to& \exp \,
(\mp i L_y \xi_1 / 4l^2)\nonumber\\
&\times& \Psi^{(a=2, m +1/2 \pm 1/2)}({\bf r}_1, \ldots,
{\bf r}_{N_e}; {\bf R}_1, {\bf R}_2), \nonumber\\
{\bf R}_1 \to {\bf R}_1 \pm L_x {\bf {\hat x}} : \Psi^{(a=4,m)}({\bf
r}_1, \ldots, {\bf r}_{N_e}; {\bf R}_1, {\bf R}_2 ) &=& (-1)^{N_e/2}
e^{i \pi m (1/2 \pm 1/2)} \nonumber\\
&\times& \Psi^{(a=3,m)}({\bf r}_1, \ldots, {\bf r}_{N_e}; {\bf R}_1,
{\bf R}_2 ). \label{EQ:MRtransQH}
\end{eqnarray}
We see that, unlike for the Laughlin states, these states are not
eigenstates with respect to $w_j \rightarrow w_j \pm 1$
transformations. However, the feature  that this transformation does
not change $m$ persists. For the translations $w_j \rightarrow w_j
\pm \tau$, there is a phase factor due to the gauge transformation
${\bf A} \rightarrow {\bf A} - BL_y{\bf {\hat x}}$. Comparing the
gauge transformation phase factors of Eq.~(\ref{EQ:transQH2}) to
that of Eq.~(\ref{EQ:MRtransQH}) confirms  that the quasiholes in
Eq.~(\ref{EQ:MRtransQH}) have charge $e/4$.

Eq.~(\ref{EQ:halfTrans}) left the following two cases unmentioned:
\begin{eqnarray}
w_1 \to w_1 \pm 1 : f^{(a=2,m)}(z_1, \ldots, z_{N_e}; w_1, w_2)
&\to& (-1)^{N_e/2}
e^{-i \pi (1/4 \mp 1/4)} e^{i \pi (m - 1) (1/2 \mp 1/2)} \nonumber\\
&\times& {\tilde f}^{(m)}(z_1, \ldots, z_{N_e}; w_1,
w_2),\nonumber\\
w_1 \to w_1 \pm \tau : f^{(a=4,m)}(z_1, \ldots, z_{N_e}; w_1, w_2)
&\to& e^{\mp i\pi/4} \exp [-i \pi N_s (\pm w_1/2 + \tau/4)] \nonumber\\
&\times& {\tilde f}^{(m + 1/2 \pm 1/2)}(z_1, \ldots,
z_{N_e};w_1,w_2), \label{EQ:halfTrans1}
\end{eqnarray}
where
\begin{eqnarray}
{\tilde f}^{(m)}(z_1, \ldots, z_{N_e}; w_1, w_2) &=& {\tilde
F}_{{\rm cm}}^{(m)} \left( \sum_i z_i + \frac{w_1 + w_2}{4} \right)
[\vartheta_1(w_{12}/2)]^{1/2} {\rm Pf} \left(-{\tilde
M}^{\alpha =(1/2, 1/2)}_{ij}\right) \nonumber\\
&\times& \prod_{i<j} [\vartheta_1 (z_i - z_j)]^2.
\label{EQ:wrongHalfQH}
\end{eqnarray}
The tilde mark is  placed above $f$ in Eq.~(\ref{EQ:wrongHalfQH})
because, in some sense, it  does {\it not} qualify as a Moore-Read
state wave function.  When the two quasiholes are merged in
Eq.~(\ref{EQ:wrongHalfQH}), the wave function simply vanishes; 
one does not get one of the wave functions of Eq.~(\ref{EQ:fullQH}).
This does not mean that Eq.~(\ref{EQ:wrongHalfQH})
necessarily gives higher energy than Eq.~(\ref{EQ:halfQH1}).
However, note that Eq.~(\ref{EQ:fullQH}) differs from the ground
states only by one quantum flux; Eq.~(\ref{EQ:wrongHalfQH}) does not
have a corresponding ground state in this sense, hence its disqualification.
One can further regard having two $e/4$ quasiholes as differing from
having a quasihole-quasiparticle pair merely by one quantum flux.
This leads to the conclusion that if one creates a quasihole-quasiparticle
pair out of one of $a$=2 (or $a$=4) ground states, translate the
quasihole around the generator in $x$ (or $y$) direction, and annihilate
the pair, one does {\it not} return to a ground state; this means
that the monodromy process can actually excite the system.

To find out what kind of excitation do we have here, we need to examine
the vanishing of Eq.~(\ref{EQ:wrongHalfQH}) as two quasiholes are brought
together. An examination into Eq.~(\ref{EQ:frac1}) shows that the vanishing
is not due to the Pfaffian part of Eq.~(\ref{EQ:wrongHalfQH}). Indeed
\begin{equation}
\lim_{w_2 \rightarrow w_1}{\tilde M}^{\alpha =(1/2, 1/2)}_{ij} =
-\frac{\vartheta_1 (z_i - w_1) \vartheta_1 (z_j -
w_1)}{\vartheta'_1(0)} \left[ \frac{\vartheta'_1(z_i -
z_j)}{\vartheta_1(z_i - z_j)} - \left(\frac{\vartheta_1^{'} (z_i -
w_1)}{\vartheta_1 (z_i - w_1)} - \frac{\vartheta'_1 (z_j -
w_1)}{\vartheta_1 (z_j - w_1)} \right)\right]. \label{EQ:frac11}
\end{equation}
The vanishing of Eq.~(\ref{EQ:wrongHalfQH}) is therefore solely due
to the $[\vartheta_1(w_{12}/2)]^{1/2}$. This term originates from
the correlator of two quasiholes in the $\alpha =(1/2, 1/2)^T$
sector. This is the sector with the periodic boundary conditions
around both generators for $\psi$ fields:
\begin{equation}
\langle \sigma(w_1) \sigma(w_2)\rangle_{\alpha =(1/2, 1/2)}
[\vartheta_1 (w_{12})]^{1/8} \propto [\vartheta_1(w_{12}/2)]^{1/2}.
\end{equation}
The spin structure is therefore correlated with the  internal state
(fusion channel) of the two quasiholes. Since for $w \rightarrow 0$
\begin{equation}
\langle \sigma(w) \sigma(0)\rangle_{\alpha =(1/2, 1/2)} \sim
w^{3/8},
\end{equation}
whereas for $\alpha = (1/2, 0)^T$, $(0, 0)^T$, and $(0, 1/2)^T$
\begin{equation}
\langle \sigma(w) \sigma(0)\rangle_\alpha \sim w^{-1/8},
\end{equation}
the chiral Ising model operator product expansion
\begin{equation}
\sigma(w) \sigma(0) \sim \frac{{\mathbb I}}{w^{1/8}} + {\rm const.}
w^{3/8} \psi(0)
\label{EQ:IsingOPE}
\end{equation}
tells us that for $\alpha =(1/2, 1/2)^T$ two $\sigma$'s fuse into
$\psi$, whereas  for $\alpha = (1/2, 0)^T$, $(0, 0)^T$, and $(0,
1/2)^T$  they fuse into ${\mathbb I}$. We see that  our wave
function approach makes manifest  the observation of  Oshikawa {\it
et al.} that, after the translation of a quasihole around a
generator,  the system may refuse to  go back into a ground state
because of a change in the $\sigma\times \sigma$ fusion channel
\cite{masaki}.  This change was explained  by Oshikawa {\it et al.}
by using the branch cut argument of Ivanov \cite{ivanov} and Stern
{\it et al}. \cite{stern}

From the preceding arguments, one can explain this change of the
fusion channel in the terms of the physical fusion in the manner
discussed by Stone and Chung for the two-dimensional $p+ip$
superconductor \cite{stone-chung}. In the ground states, two
$\sigma$'s fused into ${\mathbb I}$ and the total fermion number on
torus was even, enabling all fermions to be paired up. However, once
the translation of one quasihole rolls the system over to the
$\alpha =(1/2, 1/2)^T$ sector, the $\sigma$'s fused into a $\psi$,
which, due to the conservation of total fermion number, is possible
only if there is depairing of one of fermion pairs. Therefore, with
this change of fusion channel, 
the system is left in an excited state.

This superconductor analogy also indicates that the parity of
electron number in the ground state of the $\alpha =(1/2, 1/2)^T$
sector should be different from that of other sector. In this
picture, after the roll-over to the $\alpha =(1/2, 1/2)^T$ sector,
there is a Bogoliubov quasiparticle excitation in the system. Now if
superconductor with one Bogoliubov quasiparticle excitation has $N$
electrons, this is equivalent to having a superposition of one hole
excitation on a ground state with $N+1$ electrons and one particle
excitation on a ground state with $N-1$ electrons. Thus the change
of fusion channel has to be accompanied by the change in the parity
of the electron number in the ground state.

As previously commented, in discussing topological features, creating a
quasiparticle-quasihole pair is equivalent to splitting a charge $e/2$
quasihole into two $e/4$ quasiholes; the only difference between two
cases is that there is one more flux quantum for the latter case. In
their paper, Oshikawa {\it et al.} defined ${\mathcal T}_{x,y}$ to
be a process in which creates a quasiparticle-quasihole pair is
created and then the quasiparticle is dragged around the generator
of the torus in $x$ (or $y$) direction to wrap around the system
before it is pair-annihilated with the quasihole \cite{masaki}. So
one can make following correspondences between the analytic
continuation of theta functions considered in this paper and the
processes defined by Oshikawa {\it et al.}:
\begin{eqnarray}
w_1 &\rightarrow& w_1 + 1 \Leftrightarrow {\mathcal T}_x,\nonumber\\
w_1 &\rightarrow& w_1 + \tau \Leftrightarrow {\mathcal T}_y.
\label{EQ:OpCorr}
\end{eqnarray}
It will be shown in the appendix that the ground states labeled by
$a$ and $m$ in this paper are eigenstates of ${\mathcal T}_x^{-2}$
and ${\mathcal T}_y^4$. Labeling eigenvalues for these operators
divided by the gauge transformation factor as $f_y$ and $f'_x$
respectively, we obtain  the following correspondences between the
wave functions of this paper and the state vectors of Oshikawa {\it
et al.}:
\begin{eqnarray}
\Psi^{(a=2,m=0)} &\leftrightarrow& |f_y = i, \, f'_x = 1\rangle
\,\,\,\,\,\,\,\,\,\,\,\,\,\,\,\,\,\,\,\,\,\,\,\,\,\,\,\,\,\,\,\,\,\,\,\,
\Psi^{(a=2,m=1)}\leftrightarrow |f_y = -i, \, f'_x = 1\rangle, \nonumber\\
\Psi^{(a=3,m=0)} &\leftrightarrow& |f_y = 1, \, f'_x = 1\rangle
\,\,\,\,\,\,\,\,\,\,\,\,\,\,\,\,\,\,\,\,\,\,\,\,\,\,\,\,\,\,\,\,\,\,\,
\Psi^{(a=3,m=1)}\leftrightarrow |f_y = -1, \, f'_x = 1\rangle, \nonumber\\
(-1)^{N_e/2}\Psi^{(a=4,m=0)} &\leftrightarrow& |f_y = 1, \, f'_x =
-1\rangle \,\,\,\,\,\,\,\, (-1)^{N_e/2}\Psi^{(a=4,m=1)}
\leftrightarrow|f_y = -1, \, f'_x = -1\rangle.\nonumber\\
\label{EQ:WFStates}
\end{eqnarray}
Adapting the continuation formulas of Eq.~(\ref{EQ:halfTrans1}) to
the correspondence made in Eq.~(\ref{EQ:WFStates}), we find the
following topological actions
\begin{eqnarray}
&\,&|f_y = 1, \, f'_x = 1\rangle \,\,\,\,\,\,\,\,\,\,\,\,\,\,\,\,\,
{\mathcal T}_x |f_y = 1, \, f'_x = 1\rangle = |f_y = 1, \,f'_x
=-1\rangle, \nonumber\\
{\mathcal T}_y |f_y = 1, \, f'_x = 1\rangle
&=& |f_y = -i, \, f'_x = 1\rangle \,\,\,\,\,\,\, {\mathcal T}_x
{\mathcal T}_y |f_y = 1, \,f'_x=1\rangle = 0, \nonumber\\
{\mathcal T}_y^2 |f_y = 1, \, f'_x = 1\rangle &=& |f_y = -1, \, f'_x
= 1\rangle \,\,\,\,\, {\mathcal T}_x {\mathcal T}_y^2 |f_y = 1, \,
f'_x =1\rangle =|f_y = -1, \, f'_x = -1\rangle, \nonumber\\
{\mathcal T}_y^3 |f_y = 1, \, f'_x = 1\rangle &=& |f_y = i, \, f'_x
= 1\rangle \,\,\,\,\,\,\,\,\,\,\, {\mathcal T}_x {\mathcal T}_y^3
|f_y = 1, \, f'_x = 1\rangle = 0.\nonumber\\
\label{EQ:StateTrans}
\end{eqnarray}
There are  exactly the transformation formul\ae\ of  Oshikawa {\it
et al.} \cite{masaki} Note, however, that for our  formula, there is
a provisio that at after all the quasihole translation has been
carried out, the two quasiholes are to be merged . As in
Eq.~(\ref{EQ:transQH2}), we have dropped the gauge transformation
phase factor.

Oshikawa {\it et al.} considered bases diagonalized with respect to
either ${\mathcal T}_x$ or ${\mathcal T}_y$ \cite{masaki}. These two
bases are related by the modular $S$-matrix. In either basis, it can
be shown that the transformation formul\ae\ of this subsection
follow the Verlinde formula on diagonalization of fusion numbers by
the $S$-matrix \cite{francesco,verlinde}.

\subsection{Odd-spin structure sector}

We now need to consider the {\it odd} spin structure in detail. On a
torus, this is the case where the chiral Majorana fermion has
periodic boundary conditions around the both generators. The action
of a chiral Majorana fermion field $\psi$ is
\begin{equation}
S = \frac{1}{2\pi} \int d^2z \, \psi {\bar \partial} \psi.
\label{EQ:FAction}
\end{equation}
In performing the path integral, it is necessary to expand $\psi$ in
terms of the  normalized $c$-number eigenmodes  $\psi_{nm}(z,{\bar
z})$ of ${\bar \partial}$,. These must be  doubly periodic, and so
are
\begin{equation}
\psi_{nm}(z,{\bar z}) = \frac 1{\sqrt{{\rm Im\,} \tau} }\exp
\frac{\pi}{{\rm Im\,}\tau} [ n (\tau {\bar z} - {\bar \tau} z) + m
(z - {\bar z})]. \label{EQ:modeOdd}
\end{equation}
 The resulting mode expansion is
\begin{equation}
\psi(z, {\bar z}) = \sum_{n,m \in \mathbb{Z}} a_{nm} \psi_{mn}(z,
{\bar z}) =  a_{00} + \sideset{}{'}\sum_{n,m \geq 0} [a_{nm}
\psi_{nm}(z,{\bar z}) + a_{-n,-m} \psi_{-n,-m}(z,{\bar z})],
\label{EQ:MajoranaEx}
\end{equation}
where $a_n$'s are Grassman variables with $a_{-n,-m} \equiv
a_{nm}^*$ and $\sum'_{nm}$ is a summation over non-negative integers
that excludes the $n=m=0$ term. It should be noted that since $\psi$
is holomorphic only as a result of the equation of motion, one
cannot take it to be holomorphic in performing the path integral.

When the mode expansion Eq.~(\ref{EQ:MajoranaEx}) is inserted into
the action of Eq.~(\ref{EQ:FAction}), we obtain
\begin{equation}
S = \frac 1{2\pi}  \sideset{}{'}\sum_{n,m \geq 0} (n\tau - m)
a_{-n,-m}a_{nm}.
\end{equation}
The Grassmann variable $a_{00}$ does not appear in this action.
However, $a_{00}$ does appear in the integration measure \be d[\psi]
= \prod_{n,m \in \mathbb{Z}} da_{nm}. \ee This leads to the odd
spin-sector partition function vanishing
\cite{francesco,francesco_paper,ginsparg}:
\begin{equation}
\int da_{00} \left(\sideset{}{'}\prod_{n,m \geq 0} da_{nm}
da_{-n,-m} \right)\exp(-S) =0.
\end{equation}
As in the $\sum'_{nm}$ of Eq.~(\ref{EQ:MajoranaEx}), $\prod_{nm}'$
is a product of terms with non-negative integers $n,m$ that excludes
 the $n=m=0$ term. Now  $\int a\,da=1$ and $\int da =0$ for
any Grassmann variable $a$, and so the integration $\int F(a)
\prod_{n,m} da_{nm} $ yields  zero unless each $a_{nm}$ appears
exactly once in the integrand $F(a)$, and there is no $a_{00}$ in
the partition function integrand. For the same reason, any
correlator of an even number of $\psi$'s vanishes.

Correlators with {\it odd} number of $\psi$'s can be non-zero.
Consider the simplest case:
\begin{eqnarray}
\langle \psi(z) \rangle \propto \int  d[\psi] \psi(z) \exp(-S) &\propto&
\int a_{00} \,da_{00} \left(\sideset{}{'}\prod_{n,m \geq 0} da_{nm}
da_{-n,-m}\right) \exp(-S)\nonumber\\
&=&  \int \left(\sideset{}{'}\prod_{n,m} da_{nm} da_{-n,-m}\right)
\exp(-S) \neq 0.
\end{eqnarray}
Thus $\langle \psi(z) \rangle$ is a {\it nonzero\/} constant
\cite{francesco,francesco_paper}. The next simplest case is the
three-point correlators, where three $\psi$ fields take turns in
occupying the zero mode:
\begin{eqnarray}
\langle \psi(z_1) \psi(z_2) \psi(z_3)\rangle &\propto& \int
d[\psi] \psi(z_1) \psi(z_2) \psi(z_3) \exp(-S)\nonumber\\
&\propto&  \int \left(\sideset{}{'}\prod_{n,m \geq 0} da_{nm}
da_{-n,-m}\right) \sideset{}{'}\sum_{i,j \geq 0} a_{ij}
\psi_{ij}(z_1) \sideset{}{'}\sum_{k,l \geq 0} a_{kl} \psi_{kl}(z_2)
\exp(-S) \nonumber\\
&+&  \int \left(\sideset{}{'}\prod_{n,m \geq 0} da_{nm}
da_{-n,-m}\right) \sideset{}{'}\sum_{i,j \geq 0} a_{ij}
\psi_{ij}(z_2) \sideset{}{'}\sum_{k,l \geq 0} a_{kl} \psi_{kl}(z_3)
\exp(-S) \nonumber\\
&+&  \int \left(\sideset{}{'}\prod_{n,m \geq 0} da_{nm}
da_{-n,-m}\right) \sideset{}{'}\sum_{i,j \geq 0} a_{ij}
\psi_{ij}(z_3) \sideset{}{'}\sum_{k,l \geq 0} a_{kl} \psi_{kl}(z_1)
\exp(-S) \nonumber\\
&=& g_{++}({\bf r}_1 - {\bf r}_2) + g_{++}({\bf r}_2 - {\bf r}_3) +
g_{++}({\bf r}_3 - {\bf r}_1), \label{EQ:threeCorrel}
\end{eqnarray}
where \cite{read-rezayi_paired,read-green}
\begin{equation}
g_{++}({\bf r}) = \frac{\vartheta'_1(z)}{\vartheta_1(z)} +
\frac{2\pi i y}{L_y}. \label{EQ:oddPair2}
\end{equation}
is a modified Green function satisfying
\begin{equation}
{\bar \partial} g_{++}({\bf r}) = \pi \delta^{(2)}({\bf r}) -
\frac{\pi}{{\rm Im} \tau}.
\end{equation}
This Green function  contains $y=(z+\bar z)/2$, and is  not
holomorphic. The $\bar z$'s, however,  cancel in sum of {\it
three\/} Green functions   appearing in Eq.~(\ref{EQ:threeCorrel}).
The $3$-point correlator is, therefore, holomorphic.

In the odd spin structure, and for $N_e$ odd, the general
$N_e$-point chiral Majorana fermion correlator  is
\begin{eqnarray}
\langle \psi(z_1) \ldots \psi(z_{N_e}) \rangle &\propto&
\Psi_{++}(z_1, \ldots, z_{N_e})\nonumber\\ &=&
\frac{1}{2^{(N_e-1)/2}[(N_e-1)/2]!} \sum_{P \in S_{N_e}} {\rm
sgn}(P) \prod_{i=1}^{(N_e-1)/2} g_{++}({\bf r}_{P(2i-1)} - {\bf
r}_{P(2i)}).\nonumber\\
\label{EQ:oddPair}
\end{eqnarray}
Here $P$ runs over all permutations of $N_e$ objects. Again, at
first sight, this equation  looks neither chiral nor holomorphic.
However, Read and his coworkers \cite{read-rezayi_paired,read-green}
asserted that the cancelation in Eq.~(\ref{EQ:threeCorrel})
generalizes to
\begin{equation}
\Psi_{++}(z_1, \ldots, z_{N_e}) =
\frac{1}{2^{(N_e-1)/2}[(N_e-1)/2]!} \sum_{P \in S_{N_e}} {\rm
sgn}(P) \prod_{i=1}^{(N_e-1)/2} \frac{\vartheta'_1(z_{P(2i-1)} -
z_{P(2i)})}{\vartheta_1(z_{P(2i-1)} - z_{P(2i)})}.
\label{EQ:oddPair1}
\end{equation}
This rather non-obvious cancelation  of the $\bar z$'s will be
proved in the appendix.

Eq.~({\ref{EQ:oddPair}) was first constructed by Read and Green in
their work on spinless $p+ip$ superconductor \cite{read-green}. They
have pointed out that the existence of the ${\bf k} = 0$ mode, which
is an almost exact analog of the zero mode discussed here, requires
the number of electrons to be odd, as long as it is energetically
favorable to have this ${\bf k} = 0$ mode occupied.

We emphasize again that these periodic and anti-periodic ``boundary
conditions''  for the $\psi(z)$'s  are {\it not} the boundary
conditions of the physical  electrons. The physical  boundary
conditions are those of Eq.~(\ref{EQ:BC1}) and (\ref{EQ:BC2}). The
boundary conditions for the $\psi(z)$'s  affect only the manner of
BCS pairing. The best  way of regarding Eq.~(\ref{EQ:oddPair}) is
that this is the correct replacement for the Pfaffian when the
number of electrons  is odd. The center-of-mass wave functions
ensure that all Moore-Read states on a torus  have the same physical
boundary conditions under translation of electrons around
generators, regardless of whether the number of electrons is even or
odd. This means that ${\tilde F}^{(m)}_{{\rm cm}}$ of
Eq.~(\ref{EQ:cmWFPfaffOdd}) is  the center-of-mass wave function
when the number of electrons on the torus is odd. The holomorphic
part of the ground state wave function in this case is therefore
\begin{equation}
f^{(m)}_{odd}(z_1, \ldots, z_{N_e}) = {\tilde F}^{(m)}_{{\rm cm}}
\left( \sum_i z_i \right) \Psi_{++}(z_1, \ldots, z_{N_e})
\prod_{i<j} [\vartheta_1 (z_i - z_j)]^2.
\end{equation}
Obtaining the wave functions with one charge $e/2$ quasihole with
one quantum flux is now straightforward. It is in essentially same
form as Eq.~(\ref{EQ:fullQH}):
\begin{equation}
f^{(m)}_{odd}(z_1, \ldots, z_{N_e}; w) = {\tilde F}^{(m)}_{{\rm cm}}
\left( \sum_i z_i + \frac{{w}}{2} \right) \Psi_{++}(z_1, \ldots,
z_{N_e}) \prod_i \vartheta_1 (z_i - w) \prod_{i<j} [\vartheta_1 (z_i
- z_j)]^2.
\label{EQ:fullQHOdd}
\end{equation}

How one do we  fractionalize the quasihole of
Eq.~(\ref{EQ:fullQHOdd}) to obtain two charge $e/4$ quasiholes?  The
first formula of Eq.~(\ref{EQ:IsingLim}), being independent of the
parity of number of $\psi$'s, still holds. Also note if we define
${\tilde M}^{odd}_{ij} = -{\tilde M}^{\alpha=(1/2,1/2)}_{ij}$ then
\begin{equation}
\lim_{z_i \to z_j} \frac{(z_i - z_j)\vartheta'_1(0) {\tilde
M}^{odd}_{ij}}{[\vartheta_1(z_i - w_1)\vartheta_1(z_i -
w_2)\vartheta_1(z_j - w_1)\vartheta_1(z_j - w_2)]^{1/2}} = 1.
\end{equation}
Therefore, in the odd spin structure for $N_e$ odd, the chiral Ising
field correlator of Eq.~(\ref{EQ:manyIsingCorrel}) becomes
\begin{eqnarray}
\langle \psi(z_1) \cdots \psi(z_{N_e}) \sigma(w_1) \sigma(w_2)
\rangle_{\alpha = (1/2,1/2)} &=& \prod_{i=1}^{N_e}[\vartheta_1(z_i -
w_1)\vartheta_1(z_i-w_2)]^{-1/2} \nonumber\\ &\times& \sum_{P \in
S_{N_e}} {\rm sgn}(P) [\vartheta_1 (z_{P(N_e)} - w_1)\vartheta_1
(z_{P(N_e)} - w_2)]^{1/2} \nonumber\\
&\times& \langle \psi(z_{P(N_e)}) \sigma(w_1) \sigma(w_2)
\rangle_{\alpha = (1/2,1/2)}\nonumber\\
&\times& \prod_{j=1}^{(N_e-1)/2} \left[\vartheta'_1(0) {\tilde
M}^{odd}_{P(2j-1),P(2j)}\right].
\label{EQ:manyIsingCorrel1}
\end{eqnarray}
Note that one now needs to calculate $\langle \psi \sigma \sigma
\rangle_{\alpha = (1/2,1/2)}$ rather than $\langle \sigma
\sigma \rangle_{\alpha = (1/2,1/2)}$. 
This leads to the conclusion that with two $e/4$ quasiholes at $w_1$
and $w_2$, Eq.~(\ref{EQ:fullQHOdd}) should be modified into
\begin{eqnarray}
{\tilde \Psi}_{++}(z_1, \ldots, z_{N_e}; w_1, w_2) = &{\rm const.}&
\sum_{P \in S_{N_e}} {\rm sgn}(P) [\vartheta_1 (w_1 -
w_2)]^{1/8}\nonumber\\
&\times&[\vartheta_1 (z_{P(N_e)} - w_1)]^{1/2} [\vartheta_1
(z_{P(N_e)} - w_2)]^{1/2}\nonumber\\
&\times& \langle \psi(z_{P(N_e)}) \sigma (w_1) \sigma (w_2)
\rangle_{\alpha = (1/2,1/2)} \prod_{i=1}^{(N_e-1)/2}
\left[{\tilde M}^{odd}_{P(2i-1),P(2i)}\right].\nonumber\\
\end{eqnarray}

We compute the   necessary  Ising  correlator  in the appendix. We
find that
\begin{eqnarray}
\langle \psi(z) \sigma(w_1) \sigma(w_2) \rangle_{\alpha = (1/2,1/2)}
&\propto&
\left[\frac{1}{\vartheta_1(w_{12})}\right]^{1/8}\nonumber\\
&\times& \left[\vartheta_1^{'}(w_{12}/2) +
\frac{1}{2}\vartheta_1(w_{12}/2)\left(\frac{\vartheta_1^{'}(z-
w_1)}{\vartheta_1(z-w_1)} -
\frac{\vartheta_1^{'}(z-w_2)}{\vartheta_1(z-w_2)}\right)\right]^{1/
2}.\nonumber\\
\end{eqnarray}
Consequently, ignoring a multiplicative constant, we have
\begin{equation}
{\tilde \Psi}_{++}(z_1, \ldots, z_{N_e}; w_1, w_2) = \sum_{P \in
S_{N_e}} {\rm sgn}(P) \, [h(z_{P(N_e)};w_1,w_2)]^{1/2}
\prod_{i=1}^{(N_e-1)/2} \left[{\tilde
M}^{odd}_{P(2i-1),P(2i)}\right], \label{EQ:OddFrac}
\end{equation}
where
\begin{eqnarray}
h(z;w_1,w_2) &=& \vartheta_1^{'}(w_{12}/2)
\vartheta_1 (z - w_1) \vartheta_1 (z - w_2) \nonumber\\
&+& \frac{1}{2} \vartheta_1(w_{12}/2)\left(\vartheta_1^{'}(z -
w_1)\vartheta_1(z-w_2) - \vartheta_1^{'}(z -
w_2)\vartheta_1(z-w_1)\right). \label{EQ:unpaired}
\end{eqnarray}
The holomorphic part of the wave function with two charge $e/4$
quasiholes is
\begin{eqnarray}
f^{(m)}_{odd}(z_1, \ldots, z_{N_e}; w_1, w_2) &=& {\tilde
F}^{(m)}_{{\rm cm}} \left( \sum_i z_i + \frac{w_1 + w_2}{4} \right)
{\tilde \Psi}_{++}(z_1, \ldots, z_{N_e}; w_1, w_2) \prod_{i<j}
[\vartheta_1 (z_i - z_j)]^2\nonumber\\
&=& {\tilde F}^{(m)}_{{\rm cm}}\left( \sum_i z_i + \frac{w_1 +
w_2}{4} \right) \prod_{i<j} [\vartheta_1 (z_i - z_j)]^2\nonumber\\
&\times& \sum_{P \in S_{N_e}} {\rm sgn}(P) \,
[h(z_{P(N_e)};w_1,w_2)]^{1/2} \prod_{k=1}^{(N_e-1)/2} \left[{\tilde
M}^{odd}_{P(2k-1),P(2k)}\right]. \label{EQ:halfQHOdd}
\end{eqnarray}

It is not obvious that this wave function Eq.~(\ref{EQ:halfQHOdd})
is analytic in electron coordinates. We need to show  that
$[h(z;w_1,w_2)]^{1/2}$ is an analytic function of $z$. The proof of
the analyticity of $[h(z;w_1,w_2)]^{1/2}$ can be presented in two
steps. First, note that from
\begin{eqnarray}
h(z+1;w_1,w_2) &=& h(z;w_1,w_2) \nonumber\\
h(z+\tau;w_1,w_2) &=& \exp[-2 \pi i  (2z -w_1 - w_2 + \tau)]
h(z;w_1,w_2)
\end{eqnarray}
the $z$ of $h(z;w_1,w_2)$ should have 2 zeros in the first principal
region. Second, note that both $h(z;w_1,w_2)$ and $\partial_z
h(z;w_1,w_2)$ vanish at $z = (w_1 + w_2)/2$. This shows that there
is actually a {\it double\/} zero at $z = (w_1 + w_2)/2$ and this is
enough to ensure that $[h(z;w_1,w_2)]^{1/2}$ is indeed analytic in
$z$.

We should also   verify  that  the odd-spin-sector wave function
with two $e/4$ quasiholes goes  smoothly  into the corresponding
wave function with one $e/2$ quasihole (up to a overall
multiplicative constant) as the $e/4$ quasiholes merge. In other
words, in the limit that $w_1,w_2 \rightarrow w$, does   ${\tilde
\Psi}_{++}(z_1, \ldots, z_{N_e}; w_1, w_2)$ of
Eq.~(\ref{EQ:OddFrac}) become  proportional to $g_{++}(z_1, \ldots,
z_{N_e}) \prod_i \vartheta_1 (z_i - w)$? From Eq.~(\ref{EQ:frac11}),
\begin{eqnarray}
\lim_{w_1,w_2 \to w}{\tilde \Psi}_{++}(z_1, \ldots, z_{N_e}; w_1,
w_2) &=& [\vartheta'_1(0)]^{1/2} \prod_i \vartheta_1 (z_i - w)
\sum_{P \in S_{N_e}} {\rm sgn}(P)
\prod_{i=1}^{(N_e-1)/2}\frac{1}{\vartheta'_1(0)}\nonumber\\
&\times& \left[ \frac{\vartheta'_1(z_{P(2i-1)} -
z_{P(2i)})}{\vartheta_1(z_{P(2i-1)} - z_{P(2i)})} -
\left(\frac{\vartheta'_1 (z_{P(2i-1)} - w)}{\vartheta_1 (z_{P(2i-1)}
- w)} - \frac{\vartheta'_1 (z_{P(2i)} - w)}{\vartheta_1 (z_{P(2i)} -
w)}
\right)\right]\nonumber\\
&=& {\rm const.}\prod_i \vartheta_1 (z_i - w) \sum_{P \in S_{N_e}}
{\rm sgn}(P) \prod_{i=1}^{(N_e-1)/2} \frac{\vartheta'_1(z_{P(2i-1)}
- z_{P(2i)})}{\vartheta_1(z_{P(2i-1)} -
z_{P(2i)})}\nonumber\\
&=& {\rm const.} \Psi_{++}(z_1, \ldots, z_{N_e}) \prod_i \vartheta_1
(z_i - w). \label{EQ:oddFusion}
\end{eqnarray}
The cancelation of the  $w$-dependent terms other than $\prod_i
\vartheta_1 (z_i - w)$ occurs for exactly same reason as the one in
Eq.~(\ref{EQ:oddPair1}). This cancelation will be proved  in the
appendix.

From
\begin{eqnarray}
h(z;w_1 \pm 1,w_2) &=& \pm \vartheta_2^{'}(w_{12}/2) \vartheta_1 (z
- w_1) \vartheta_1 (z - w_2) \nonumber\\
&\pm& \frac{1}{2} \vartheta_2(w_{12}/2)\left(\vartheta'_1(z -
w_1)\vartheta_1(z-w_2) -
\vartheta'_1(z - w_2)\vartheta_1(z-w_1)\right),\nonumber\\
h(z;w_1 \pm \tau,w_2) &=& \pm i e^{-i\pi\tau/4} e^{\mp i \pi
w_{12}/2} e^{-i\pi\tau} e^{\pm 2 \pi i z}[\vartheta'_4(w_{12}/2)
\vartheta_1 (z - w_1) \vartheta_1
(z - w_2) \nonumber\\
&+& \frac{1}{2} \vartheta_4(w_{12}/2)\left(\vartheta'_1(z -
w_1)\vartheta_1(z-w_2) - \vartheta'_1(z -
w_2)\vartheta_1(z-w_1)\right)], \label{EQ:oddBlock}
\end{eqnarray}
it is clear that after after one quasihole is translated around a
generator, the wave function vanishes when two quasiholes are
brought together.

As  in the even spin structure case, this vanishing can be
attributed to the change in the fusion channel of two $\sigma$'s.
This is so since
\begin{equation}
\lim_{w_1 \to w_2} \langle \psi(z) \sigma(w_1) \sigma(w_2)
\rangle_{\alpha = (1/2,1/2)} \sim w_{12}^{-1/8},
\end{equation}
but after the translation $w_1 \to w_1 \pm 1$ or  $w_1 \to w_1 \pm
\tau$, Eq.~(\ref{EQ:oddBlock}) tells us that the same  correlator
vanishes as $w_{12}^{3/8}$ when $w_1 \to w_2$. From the chiral Ising
operator product expansion, Eq.~(\ref{EQ:IsingOPE}), one can see two
$\sigma$'s fuse to ${\mathbb I}$ in $\langle \psi \sigma \sigma
\rangle_{\alpha = (1/2,1/2)}$, but after the $w_1 \to w_1 \pm 1$ or
$w_1 \to w_1 \pm \tau$ translation, they  fuse to $\psi$. The
argument made in the last subsection that this  change in the fusion
channel is accompanied by the change in the parity of the number of
the electrons in the ground state remain valid.  Since the number of
electrons in the system does not change, a quasiparticle excitation
must have been created. Note that the different monodromy results of
the odd spin structure arises from the fact that here $\langle \psi \sigma \sigma \rangle$
plays the role of $\langle \sigma \sigma \rangle$ in the even spin
structures. We see that the different monodromy outcomes originate
from the different parities of the electron number.

From the monodromy outcomes we have obtained, the ground states of
the odd spin structure are eigenstates of both ${\mathcal T}_x$
and ${\mathcal T}_y$. However, in the even spin structures, no ground
state is simultaneously an eigenstate of both ${\mathcal T}_x$ and
${\mathcal T}_y$. The difference comes from the fact that, in the
language of the critical Ising model, the total topological charge
of the system is $\psi$ in the odd spin structure whereas it is
${\mathbb I}$ in the even spin structures. That means that the modular
$S$-matrix of the system for the odd spin structure differs from that
of the even spin structures - $S^\psi$ for the former and $S^{\mathbb I}$
for the latter \cite{kitaev_anyons,SMatrices}.

This discussion on total topological charge indicates that the odd
spin structure sector exists only because electrons that makes up a
quantum Hall system are fermions. On the other hand, a system
consisting of bosons, as in the string-net of Levin and Wen
\cite{stringNet}, one cannot have a ground state with total
topological charge of $\psi$. In such system, one would still have
`forbidden transition' due to change in the fusion channel, but this
change is not tied with the change in the parity of number of
particles in the ground state in the manner discussed here. The
change in the fusion channel can be attributed to the change in the
parity of number of particles in the ground state only when the
particles are fermions and the ground state has BCS pairing.

\section{Conclusion and Discussion}

In this paper, we constructed the two-quasihole wave functions for
the Moore-Read quantum Hall state with on a torus for both even and
odd spin structures. Conformal field theory calculations enable us
to obtain the complete  dependence of these wave functions on
quasihole coordinates. We showed that  the number of electrons in a
ground state must  be odd in the odd spin structure, and even in the
even spin structures.  By noting that the boundary conditions of
each electron remain that of Eq.~(\ref{EQ:BC1}) for all
topologically degenerate ground states, we obtained the explicit
expressions for the center-of-mass wave functions in all cases.
Analytic continuation  allowed us to obtain  the monodromy matrix
describing the effect on the space degenerate ground states of
taking quasiholes around the torus generators.  The effects are  in
agreement with those obtained by  Oshikawa {\it et al.}  In this
process, we demonstrated that otherwise anticipated  transitions are
forbidden because they  involve a  change in the fusion channel of
the quasiholes. This reflects a change in the parity of the number
of electrons that can reside in the ground state. Since the number
of electrons is conserved, the operations that might have resulted
in these forbidden operations take us out of the space of degenerate
ground states, and into the space of excited states.

Several extensions of our analysis are possible. Theta functions can
be generalized to  higher genus Riemann surface \cite{tata2}, and
these functions will  naturally be ingredients of wave functions in
such topology. Such wave functions for Laughlin states had been
studied \cite{higherG}. With these wave functions are found for the
Moore-Read state, it should be possible to extend our calculations
to the higher genus Riemann surface considered by Oshikawa {\it et
al.} \cite{masaki}. It would also be valuable to extend this
analysis to consider the next simplest non-Abelian quantum Hall
states, the Read-Rezayi parafermion states. The  number of
degenerate ground states has already been  found \cite{read-talk}.
The challenge is to work out the wave function on the compact
Riemann surface using the conformal field theory analysis, thus
extending the recent wave function construction by E.~Ardonne and
K.~Schoutens on the plane\cite{eddy_wf}.

\section{Acknowledgement}

We would like to thank E.~Fradkin, P.~Fendley, and N.~Read for
carefully reading over the earlier version of this paper and
offering comments on it. We owe special thanks to M.~Oshikawa for
the talk he gave at the Kavli Institue for Theoretical Physics
(KITP), {\it Topological Degeneracy of Non-Abelian States for
Dummies} \cite{oshikawa-talk} and M.~Levin for discussion on
$S$-matrices and topological charge. Parts of this work had been
completed at the KITP, UCSB while participating in the program on
{\it Topological Phases and Quantum Computation}. We would like to
express gratitude to the organizers, the staff, and the director for
their kind hospitality at the KITP. This research is supported by
the University of Illinois at Urbana-Champaign Research Board and
the National Science Foundation under Grant No. DMR 06-03528.

\appendix

\section{Translation eigenvalues of wave functions}

Following Eq.~(\ref{EQ:OpCorr}), ${\mathcal T}_x^{-2}$ and
${\mathcal T}_y^4$ can be regarded as implementing $\textbf{R}_1 \to
\textbf{R}_1 - L_x {\bf {\hat x}}$ twice and $\textbf{R}_1 \to
\textbf{R}_1 + L_y {\bf {\hat y}}$ four times, respectively, on
two-quasihole wave functions $\Psi^{(a,m)}({\bf r}_1, \ldots, {\bf
r}_{N_e}; {\bf R}_1, {\bf R}_2 )$. For $a=3$, repeated application
of Eq.~(\ref{EQ:halfTrans}) leads to
\begin{eqnarray}
{\mathcal T}_y^{4}\,\,\,\, : \Psi^{(a=2,m)}({\bf r}_1, \ldots, {\bf
r}_{N_e}; {\bf R}_1, {\bf R}_2 ) &\to& \exp(-i L_y \xi_1/l^2)
\Psi^{(a=2,m)}({\bf r}_1, \ldots, {\bf r}_{N_e}; {\bf R}_1, {\bf
R}_2 ),\nonumber\\
{\mathcal T}_x^{-2} : \Psi^{(a=3,m)}({\bf r}_1, \ldots, {\bf
r}_{N_e}; {\bf R}_1, {\bf R}_2 ) &\to& e^{i\pi m}
\Psi^{(a=3,m)}({\bf r}_1, \ldots, {\bf r}_{N_e}; {\bf R}_1, {\bf
R}_2
),\nonumber\\
{\mathcal T}_y^{4}\,\,\,\, : \Psi^{(a=3,m)}({\bf r}_1, \ldots, {\bf
r}_{N_e}; {\bf R}_1, {\bf R}_2 ) &\to& \exp(-i L_y \xi_1/l^2)
\Psi^{(a=3,m)}({\bf r}_1, \ldots, {\bf r}_{N_e}; {\bf R}_1, {\bf
R}_2 ),\nonumber\\
{\mathcal T}_x^{-2} : \Psi^{(a=4,m)}({\bf r}_1, \ldots, {\bf
r}_{N_e}; {\bf R}_1, {\bf R}_2 ) &\to& e^{i\pi m}
\Psi^{(a=4,m)}({\bf r}_1, \ldots, {\bf r}_{N_e}; {\bf R}_1, {\bf
R}_2 ).
\end{eqnarray}

To apply these operations on $a=2$ and $a=4$ states, one also needs
to consider the transformation of ${\tilde f}^{(m)}(z_1, \ldots,
z_{N_e}; w_1, w_2)$ of Eq.~(\ref{EQ:wrongHalfQH}), even if this does
not qualify as a Moore-Read state wave function. From the standard
theta function identities
\begin{eqnarray}
\vartheta_1(z-1/2) &=& -\vartheta_2(z),\nonumber\\
\vartheta_1(z+\tau/2) &=& i \exp[-i\pi(z+\tau/4)]\vartheta_4(z),
\label{EQ:theta1trans}
\end{eqnarray}
and the center-of-mass wave function transformation
\begin{eqnarray}
{\tilde F}^{(m)}(z-1/4) &=& F^{(a=2,m)}_{{\rm cm}}(z),\nonumber\\
{\tilde F}^{(m)}(z+\tau/4) &=& e^{i\pi(N_s-2)/4}\exp[-i\pi(z +
\tau/8)]F^{(a=4,m)}_{{\rm cm}}(z), \label{EQ:cm1trans}
\end{eqnarray}
one obtains
\begin{eqnarray}
w_1 \to w_1 - 1 : {\tilde f}^{(m)}(z_1, \ldots, z_{N_e}; w_1, w_2)
&\rightarrow& (-1)^{N_e/2} f^{(a=2,m)}(z_1, \ldots, z_{N_e}; w_1,
w_2),\nonumber\\
w_1 \to w_1 + \tau : {\tilde f}^{(m)}(z_1, \ldots, z_{N_e}; w_1,
w_2) &\rightarrow& e^{-i \pi/4} \exp[-i\pi N_s (w_1/2 + \tau/4)]
\nonumber\\
&\times& f^{(a=4,m)}(z_1, \ldots, z_{N_e}; w_1, w_2).
\label{EQ:wrongTrans}
\end{eqnarray}
Repeated application of Eqs.~(\ref{EQ:halfTrans}) and
(\ref{EQ:wrongTrans}) leads to
\begin{eqnarray}
{\mathcal T}_x^{-2} : \Psi^{(a=2,m)}({\bf r}_1, \ldots, {\bf
r}_{N_e}; {\bf R}_1, {\bf R}_2 ) &\to& -i e^{i\pi (m-1)}
\Psi^{(a=2,m)}({\bf r}_1, \ldots, {\bf r}_{N_e}; {\bf R}_1, {\bf
R}_2
),\nonumber\\
{\mathcal T}_y^{4}\,\,\,\, : \Psi^{(a=4,m)}({\bf r}_1, \ldots, {\bf
r}_{N_e}; {\bf R}_1, {\bf R}_2 ) &\to& -\exp(-i L_y \xi_1/l^2)
\Psi^{(a=4,m)}({\bf r}_1, \ldots, {\bf r}_{N_e}; {\bf R}_1, {\bf
R}_2 ).\nonumber\\
\end{eqnarray}

\section{Proof for cancelation in summation}

Consider an function $G({\bf r},{\bf r'})$ odd under exchange of
${\bf r}$ and ${\bf r'}$. Suppose this function can be expressed as
a sum of two function $g_1({\bf r},{\bf r'})$ and $g_2({\bf r},{\bf
r'})$; $g_2({\bf r},{\bf r'})$ has a further property that $g_2({\bf
r}_1,{\bf r}_2) + g_2({\bf r}_2,{\bf r}_3) + g_2({\bf r}_3,{\bf
r}_1) = 0$. Certain functions in Sec.III, such as
$g_{++}(\textbf{r})$ in Eq.~(\ref{EQ:oddPair}) and
$\tilde{M}^{odd}_{ij}$ in the limit $w_1, w_2 \rightarrow w$ as in
Eq.~(\ref{EQ:oddFusion}) (when regarded as a function of two
electron coordinates), are of this form. In order that the
assertions made in Eq.~(\ref{EQ:oddPair}) and
Eq.~(\ref{EQ:oddFusion}) hold, it is necessary to show that for $N$
odd
\begin{equation}
\sum_{P \in S_{N}} {\rm sgn}(P) \prod_{i=1}^{(N-1)/2} G({\bf
r}_{P(2i-1)},{\bf r}_{P(2)}) = \sum_{P \in S_{N}} {\rm sgn}(P)
\prod_{i=1}^{(N-1)/2} g_1({\bf r}_{P(2i-1)},{\bf r}_{P(2i)}).
\label{EQ:sumTH}
\end{equation}

One can easily see that it should be so for $N=3$ from $G({\bf
r}_1,{\bf r}_2) + G({\bf r}_2,{\bf r}_3) + G({\bf r}_3,{\bf r}_1) =
g_1({\bf r}_1,{\bf r}_2) + g_1({\bf r}_2,{\bf r}_3) + g_1({\bf
r}_3,{\bf r}_1)$, due to the property of $g_2$ mentioned above.
Suppose it holds for $N=2k+1$ where $k$ is some positive integer. If
it can be shown from this assumption that Eq.~(\ref{EQ:sumTH}) holds
for $N=2k+3$, then we have a proof by induction.

To apply induction, a new permutation $P' \in S_{2k+1}$ needs be
introduced:
\begin{eqnarray}
\sum_{P \in S_{2k + 3}} {\rm sgn}(P) \prod_{i=1}^{k+1} G({\bf
r}_{P(2i-1)},{\bf r}_{P(2i)}) &=& 2 \sum_{m<n} (-1)^{m-n-1} G({\bf
r}_m,{\bf r}_n)\nonumber\\
&\times& \sum_{P' \in S_{2k + 1}} {\rm sgn}(P') \prod_{i=1}^k G({\bf
r}_{P'(2i-1)},{\bf r}_{P'(2i)}), \label{EQ:proof1}
\end{eqnarray}
where $m,n$ are integers between 1 and $2k+3$. This $P'$ is a
permutation of integers  between 1 and $2k+3$ {\it except $m$ and
$n$}. Since the assumption had been made that Eq.~(\ref{EQ:sumTH})
holds for $N=2k+1$, Eq.~(\ref{EQ:proof1}) means
\begin{eqnarray}
\sum_{P \in S_{2k + 3}} {\rm sgn}(P) \prod_{i=1}^{k+1} G({\bf
r}_{P(2i-1)},{\bf r}_{P(2i)}) &=& 2 \sum_{m<n} (-1)^{m-n-1} G({\bf
r}_m,{\bf r}_n)\nonumber\\
&\times& \sum_{P' \in S_{2k + 1}} {\rm sgn}(P') \prod_{i=1}^k
g_1({\bf r}_{P'(2i-1)},{\bf r}_{P'(2i)}). \label{EQ:proof2}
\end{eqnarray}
Comparing Eq.~(\ref{EQ:proof1}) and Eq.~(\ref{EQ:proof2}) shows that
Eq.~(\ref{EQ:proof2}) means we can replace all but one $G$ into
$g_1$. However, this leads to the conclusion that the last remaining
$G$ can also be replaced by $g_1$:
\begin{eqnarray}
\sum_{P \in S_{2k + 3}} {\rm sgn}(P) \prod_{i=1}^{k+1} G({\bf
r}_{P(2i-1)},{\bf r}_{P(2i)}) &=& \sum_{P \in S_{2k + 3}} {\rm
sgn}(P) G({\bf r}_{P(2k+1)},{\bf r}_{P(2k+2)}) \prod_{i=2}^{k+1}
g_1({\bf r}_{P(2i-1)},{\bf r}_{P(2i)})\nonumber\\
&=& \frac{1}{3} \sum_{P \in S_{2k + 3}} {\rm sgn}(P) [G({\bf
r}_{P(2k+1)},{\bf r}_{P(2k+2)})+G({\bf r}_{P(2k+2)},{\bf
r}_{P(2k+3)})\nonumber\\ &+& G({\bf r}_{P(2k+3)},{\bf r}_{P(2k+1)})]
\prod_{i=2}^{k+1} g_1({\bf r}_{P(2i-1)},{\bf
r}_{P(2i)})\nonumber\\
&=& \frac{1}{3} \sum_{P \in S_{2k + 3}} {\rm sgn}(P) [g_1({\bf
r}_{P(2k+1)},{\bf r}_{P(2k+2)})+g_1({\bf r}_{P(2k+2)},{\bf
r}_{P(2k+3)})\nonumber\\
&+& g_1({\bf r}_{P(2k+3)},{\bf r}_{P(2k+1)})] \prod_{i=2}^{k+1}
g_1({\bf
r}_{P(2i-1)},{\bf r}_{P(2i)})\nonumber\\
&=& \sum_{P \in S_{2k + 3}} {\rm sgn}(P) \prod_{i=1}^{k+1} g_1({\bf
r}_{P(2i-1)},{\bf r}_{P(2i)}). \label{EQ:proof3}
\end{eqnarray}

\section{Critical Ising correlators on torus}

To calculate \begin{equation}\langle \psi(z) \sigma(w_1) \sigma(w_2)
\rangle_{\alpha = (1/2,1/2)},\end{equation} first note the
connection between the correlator in the chiral Ising theory and the
full Ising theory:
\begin{equation}
|\langle \psi(z) \sigma(w_1) \sigma(w_2) \rangle|^2 = \langle
\varepsilon(z, {\bar z}) \sigma(w_1, {\bar w_1}) \sigma(w_2, {\bar
w_2}) \rangle. \label{EQ:chiralIsing}
\end{equation}
So $\langle \psi \sigma \sigma \rangle$ can be computed by taking
the holomorphic part of $\langle \varepsilon \sigma \sigma \rangle$.
By the Ising model bosonization formula \cite{francesco,zuber}:
\begin{eqnarray}
\langle \varepsilon(z, {\bar z}) \sigma(w_1, {\bar w_1}) \sigma(w_2,
{\bar w_2}) \rangle^2 &=& -2 \langle
\partial \phi(z) {\bar \partial} \phi({\bar z}) \cos
\frac{\phi(w_1, {\bar w_1})}{2} \cos \frac{\phi(w_2, {\bar
w_2})}{2} \rangle \nonumber\\
&=& -\frac{1}{2} [\langle \partial \phi(z){\bar \partial} \phi({\bar
z}) \exp(i \phi(w_1,{\bar w_1})/2) \exp(-i \phi(w_2,{\bar
w_2})/2)\rangle \nonumber\\
&+& \langle \partial \phi(z){\bar \partial} \phi({\bar z}) \exp(-i
\phi(w_1,{\bar w_1})/2) \exp(i \phi(w_2,{\bar
w_2})/2)\rangle],\nonumber\\
\label{EQ:IsingBosonization}
\end{eqnarray}
where $\phi (z, {\bar z})$ is a free boson field. It should be
emphasized that there is an intricacy hidden behind
Eq.~(\ref{EQ:IsingBosonization}). Critical Ising fields on torus can
be bosonized only into a {\it compactified} free boson. A boson with
compactification radius $r=1$ should have boundary condition
\begin{equation}
\phi^{(m,m')} (z+1, {\bar z} +1) = \phi^{m,m'} (z,{\bar z}) + 2\pi
m, \,\,\,\,\, \phi^{(m,m')} (z+\tau, {\bar z} + {\bar \tau}) =
\phi^{m,m'} (z,{\bar z}) + 2\pi m', \label{EQ:compact}
\end{equation}
where $m$, $m'$ are integers. This winding comes entirely from the
zero mode, so one can express the free boson field $\phi$ as a sum
of this zero mode and ``free part'' ${\hat \phi}$ - that is, nonzero
modes:
\begin{equation}
\phi^{(m,m')} (z, {\bar z}) = \frac{\pi}{i \tau_2} [ m (\tau {\bar
z} - {\bar \tau} z) + m' (z - {\bar z})] + {\hat \phi} (z, {\bar
z}), \label{EQ:compactBoson}
\end{equation}
where $\tau = \tau_1 + i \tau_2$ with $\tau_1$,$\tau_2$ real. (Note
that in this appendix, the first principal region is a parallelogram
and not necessarily a square and nonzero $\tau_1$ is considered.)
This decomposition of the boson field leads to the following
decomposition of the action: \cite{francesco,ginsparg}
\begin{eqnarray}
S[\phi] &=& (1/8\pi) \int (\partial \phi) ({\bar \partial}
\phi) \nonumber\\
&=& S[\phi_0] + S[{\hat \phi}] - (1/4\pi) \int {\hat \phi} \Delta
\phi_0 = S[\phi_0] + S[{\hat \phi}], \label{EQ:action}
\end{eqnarray}
where $\phi_0$ refers to the zero mode terms of
Eq.~(\ref{EQ:compactBoson}). This result is due to the vanishing of
the Laplacian of $\phi_0$. This also means that the boson partition
function $Z^{bos}$ factorizes into the zero mode part $Z_0$ and the
the ``free part'' ${\hat Z}$: $Z^{bos} = Z_0 {\hat Z}$. Since in
calculating $Z_0$, all $m$, $m'$ need to be summed over, one obtains
\cite{francesco, francesco_paper, ginsparg}
\begin{equation}
Z_0 = \sum_{m,m'} Z_0^{(m,m')},\label{EQ:zeroPartition}
\end{equation}
where $Z_0^{(m,m')} = \exp [-(\pi |m\tau - m'|^2/2\tau_2)]$.

To compute a full boson correlator, one should be calculate the
correlator first for the winding sector $m$,$m'$ and then sum over
all possible winding numbers. However, this does not suffice for the
purpose here, which is to calculate $\langle \psi \sigma \sigma
\rangle_{\alpha = (1/2,1/2)}$. The question is how to extract out
the portion of the correlator that corresponds to $\alpha =
(1/2,1/2)$ sector of the Ising model.

The answer can be obtained from converting expressing theta
functions in terms of $Z_0^{(m,m')}$ of
Eq.~(\ref{EQ:zeroPartition}). Following Di Francesco {\it et al.},
\cite{francesco_paper}
\begin{eqnarray}
\left\arrowvert\vartheta \left [
\begin{array}{c}
  \frac{1}{2} \\
  \frac{1}{2} \\
\end{array}
\right ] (z\,|\,\tau)\right\arrowvert^2 &=& \sum_{n,{\bar n}} e^{i
\pi [\tau (n+1/2)^2 - {\bar \tau} ({\bar n}+1/2)^2]} e^{2 \pi i
[(n+1/2)(z + 1/2) -
({\bar n}+1/2)({\bar z}+1/2)]}\nonumber\\
&=& \left(\sum_{m \in 2 {\mathbb Z}+1} \sum_{q \in 2 {\mathbb Z}}+
\sum_{m \in 2 {\mathbb Z}} \sum_{q \in 2 {\mathbb Z}+1}\right)
(-1)^m e^ {i\pi [mq\tau_1 + (i/2)(m^2+q^2)\tau_2]} e^{\pi i [m
(z+{\bar z}) + q (z - {\bar z})]},\nonumber\\
\label{EQ:thetaBoson1}
\end{eqnarray}
where $m = n - {\bar n}$ and $q = n + {\bar n} +1$. Applying the
{\it Poisson resummation formula}
\begin{equation}
\sum_n \exp(-\pi a n^2 + bn) = \frac{1}{{\sqrt a}} \sum_k
\exp\left[-\frac{\pi}{a} (k + b/2\pi i)^2 \right]
\end{equation}
for the summation over $q$ in Eq.~(\ref{EQ:thetaBoson1}) leads to
\begin{eqnarray}
|\vartheta_1(z)|^2 &=& \frac{-1}{\sqrt {2 \tau_2}} \sum_{m \in 2
{\mathbb Z}+1} e^{-\pi m^2 \tau_2/2} e^{\pi i m (z + {\bar z})}
\left(\sum_{m'} \exp [-(\pi/2\tau_2)(m' - m\tau_1 - z + {\bar
z})^2]\right)\nonumber\\
&+& \frac{e^{\pi i (z - {\bar z}+i\tau_2/2)} }{\sqrt {2 \tau_2}}
\sum_{m \in 2 {\mathbb Z}} e^{-\pi m^2 \tau_2/2} e^{\pi i m (z+{\bar
z}+\tau_1)} \left(\sum_{m'} \exp [-(\pi/2\tau_2)(m' - m\tau_1 - z +
{\bar z} - i\tau_2)^2]\right)\nonumber\\
&=& -\frac{\exp[-\pi (z - {\bar z})^2/2 \tau_2]}{\sqrt{2\tau_2}}
\sum_{m,m'} (-1)^{(m+1)(m'+1)} Z_0^{(m,m')} \exp[i
\phi^{(m,m')}_0(z,{\bar z})],\nonumber\\
\label{EQ:thetaBoson2}
\end{eqnarray}
where $\phi^{(m,m')}_0(z,{\bar z}) = (\pi/i\tau_2)[m (\tau {\bar z}
- {\bar \tau} z) + m' (z - {\bar z})]$ as in
Eq.~(\ref{EQ:compactBoson}). (Note that the definition of the Jacobi
theta functions, Eq.~(\ref{EQ:JThetas}) is also used.)

Eq.~(\ref{EQ:thetaBoson1}) and Eq.~(\ref{EQ:thetaBoson2}) tell us
that when summing over different winding numbers $m$ and $m'$,
weighting different winding sector by the sign factor
$-(-1)^{(m+1)(m'+1)}$ would result in extracting out the ${\bf
\alpha} = (1/2,1/2)$ sector of the Ising model. It is now possible
to calculate the boson correlators of
Eq.~(\ref{EQ:IsingBosonization}), e.g.
\begin{eqnarray}
&\,&\langle \partial \phi(z){\bar \partial} \phi({\bar z}) \exp(i
\phi(w_1,{\bar w_1})/2) \exp(-i \phi(w_2,{\bar w_2})/2)
\rangle_{{\bf
\alpha} = (1/2,1/2)}\nonumber\\
&=& \frac{-1}{Z_0 {\hat Z}} \int \mathcal{D}{\hat \phi}e^{-S[{\hat
\phi}]}\sum_{m,m'}(-1)^{(m+1)(m'+1)}Z^{(m,m')}_0 \partial
\phi^{(m,m')}(z){\bar
\partial} \phi^{(m,m')}({\bar z}) \nonumber\\
&\times& \exp(i\phi^{(m,m')}(w_1,{\bar w_1})/2) \exp(-i
\phi^{(m,m')}(w_2,{\bar w_2})/2)\nonumber\\
&=&\frac{-1}{Z_0}\sum_{m,m'}(-1)^{(m+1)(m'+1)}Z^{(m,m')}_0
\exp[i\phi^{(m,m')}_0(w_{12}/2,{\bar w_{12}}/2)] \nonumber\\
&\times&[\partial \phi^{(m,m')}_0(z){\bar \partial}
\phi^{(m,m')}_0({\bar z}) \langle \exp(i {\hat \phi}(w_1,{\bar
w_1})/2) \exp(-i {\hat \phi}(w_2,{\bar
w_2})/2) \rangle\nonumber\\
&+& \partial \phi^{(m,m')}_0(z)\langle {\bar \partial} {\hat
\phi}({\bar z})  \exp(i {\hat \phi}(w_1,{\bar w_1})/2) \exp(-i {\hat
\phi}(w_2,{\bar
w_2})/2) \rangle\nonumber\\
&+& {\bar \partial} \phi^{(m,m')}_0({\bar z}) \langle \partial {\hat
\phi}(z) \exp(i {\hat \phi}(w_1,{\bar w_1})/2) \exp(-i {\hat
\phi}(w_2,{\bar w_2})/2) \rangle\nonumber\\
&+& \langle \partial {\hat \phi} (z) {\bar \partial} {\hat
\phi}({\bar z}) \exp(i {\hat \phi}(w_1,{\bar w_1})/2) \exp(-i {\hat
\phi}(w_2,{\bar w_2})/2) \rangle]. \label{EQ:a1Correl}
\end{eqnarray}
The summation over $m$ and $m'$ can be eliminated by inserting the
result of Eq.~(\ref{EQ:thetaBoson2}), together with some necessary
differentiations, into Eq.~(\ref{EQ:a1Correl}):
\begin{eqnarray}
&\,&\langle \partial \phi(z){\bar \partial} \phi({\bar z}) \exp(i
\phi(w_1,{\bar w_1})/2) \exp(-i \phi(w_2,{\bar w_2})/2)
\rangle_{{\bf
\alpha} = (1/2,1/2)}\nonumber\\
&=&\frac{\sqrt {2 \tau_2}}{Z_0}\exp [\pi (w_{12}-{\bar
w_{12}})^2/8\tau_2][\langle \exp(i {\hat \phi}(w_1,{\bar
w_1})/2) \exp(-i {\hat \phi}(w_2,{\bar w_2})/2) \rangle \nonumber\\
&\times& \{-|\vartheta'_1(w_{12}/2)|^2 + \left(\frac{\pi}{\tau_2} +
\frac{\pi^2 (w_{12} - {\bar
w_{12}})^2}{4\tau_2^2}\right)|\vartheta_1(w_{12}/2)|^2\nonumber\\
&+& \frac{\pi (w_{12} - \bar {w_{12}})}{2\tau_2}
\vartheta'_1(w_{12}/2) {\bar \vartheta_1}({\bar w_{12}}/2)-
\frac{\pi (w_{12} - {\bar w_{12}})}{2\tau_2}
\vartheta_1(w_{12}/2){\bar
\vartheta'_1}({\bar w_{12}}/2)\}\nonumber\\
&-&i\langle {\bar \partial} {\hat \phi}({\bar z})  \exp(i {\hat
\phi}(w_1,{\bar w_1})/2) \exp(-i {\hat \phi}(w_2,{\bar w_2})/2)
\rangle\nonumber\\
&\times& {\bar \vartheta_1}({\bar w_{12}}/2)
\left(\vartheta'_1(w_{12}/2) + \frac{\pi (w_{12} - {\bar
w_{12}})}{2\tau_2}\vartheta_1(w_{12}/2)\right)\nonumber\\
&-&i \langle \partial {\hat \phi}(z) \exp(i {\hat \phi}(w_1,{\bar
w_1})/2) \exp(-i {\hat \phi}(w_2,{\bar w_2})/2) \rangle \nonumber\\
&\times& \vartheta_1(w_{12}/2) \left({\bar \vartheta'_1}({\bar
w_{12}}/2)-\frac{\pi (w_{12} - {\bar w_{12}})}{2\tau_2} {\bar
\vartheta_1}({\bar w_{12}}/2)
\right)\nonumber\\
&+&\langle \partial {\hat \phi} (z) {\bar \partial} {\hat
\phi}({\bar z}) \exp(i {\hat \phi}(w_1,{\bar w_1})/2) \exp(-i {\hat
\phi}(w_2,{\bar w_2})/2) \rangle |\vartheta_1(w_{12}/2)|^2].
\label{EQ:a1Correl1}
\end{eqnarray}

Only the correlators of the ``free part'' of the boson remains to be
determined. Its propagator is
\begin{equation}
\langle {\hat \phi}(z,{\bar z}) {\hat \phi}(0,0) \rangle = -\ln
\left\arrowvert\frac{\vartheta_1(z)}{\vartheta'_1(0)}\right\arrowvert^2
- \frac{\pi(z-{\bar z})^2}{2\tau_2}. \label{EQ:freeProp}
\end{equation}
Note that this propagator is the solution to the modified Green
function satisfying
\begin{equation}
-\Delta G(z,{\bar z}) = 4\pi \delta^{(2)}(z) - \frac{4\pi}{\tau_2}.
\label{EQ:Greenftn}
\end{equation}
 From Eq.~(\ref{EQ:freeProp}), one can obtain
\begin{eqnarray}
\langle \exp(i {\hat \phi}(w_1,{\bar w_1})/2) \exp(-i {\hat
\phi}(w_2,{\bar w_2})/2) \rangle &=& \exp \left[-\frac{\pi
(w_{12}-{\bar w_{12}})^2}{8\tau_2}\right]
\left\arrowvert\frac{\vartheta'_1(0)}{\vartheta_1(w_{12})}
\right\arrowvert^{1/2}
.\nonumber\\
\label{EQ:freeCorrel1}
\end{eqnarray}
Since
\begin{eqnarray}
\langle \partial {\hat \phi}(z) \exp[i {\hat \phi}(w_1, {\bar
w_1})/2] \exp[-i {\hat \phi}(w_2, {\bar w_2})/2]\rangle &=& \exp
\left[\frac{1}{4} \langle {\hat \phi}(w_1,{\bar w_1}) {\hat
\phi}(w_2,{\bar w_2})\rangle
\right]\nonumber\\
&\times& \langle \partial {\hat \phi}(z) \exp[i ({\hat
\phi}(w_1,{\bar w_1})-{\hat \phi}(w_2,{\bar w_2}))/2] \nonumber\\
&=& \frac{i}{2} \exp \left[\frac{1}{4} \langle {\hat \phi}(w_1,{\bar
w_1}) {\hat \phi}(w_2,{\bar w_2})\rangle
\right]\nonumber\\
&\times& (\langle \partial {\hat \phi}(z) {\hat \phi}(w_1, {\bar
w_1}) \rangle - \langle \partial {\hat
\phi}(z) {\hat \phi_c}(w_2, {\bar w_2}) \rangle), \nonumber\\
\label{EQ:freeCorrel2}
\end{eqnarray}
Eq.~(\ref{EQ:freeProp}) leads to
\begin{eqnarray}
\langle \partial {\hat \phi}(z) \exp[i {\hat \phi}(w_1, {\bar
w_1})/2] \exp[-i {\hat \phi}(w_2, {\bar w_2})/2]\rangle &=&
-\frac{i}{2} \exp \left[-\frac{\pi (w_{12}-{\bar
w_{12}})^2}{8\tau_2}\right]
\left\arrowvert\frac{\vartheta'_1(0)}{\vartheta_1(w_{12})}
\right\arrowvert^{1/2}\nonumber\\
&\times& \left[\frac{\vartheta_1^{'}(z-w_1)}{\vartheta_1(z-w_1)} -
\frac{\vartheta_1^{'}(z-w_2)}{\vartheta_1(z-w_2)}-\frac{\pi (w_{12}
- {\bar w_{12}}) }{\tau_2}\right].\nonumber\\
\label{EQ:freeCorrel3}
\end{eqnarray}
Similarly
\begin{eqnarray}
\langle {\bar \partial} {\hat \phi}({\bar z}) \exp[i {\hat
\phi}(w_1, {\bar w_1})/2] \exp[-i {\hat \phi}(w_2, {\bar
w_2})/2]\rangle &=& -\frac{i}{2} \exp \left[-\frac{\pi (w_{12}-{\bar
w_{12}})^2}{8\tau_2}\right] \left\arrowvert\frac{\vartheta'_1(0)}
{\vartheta_1(w_{12})}\right\arrowvert^{1/2}\nonumber\\
&\times& \left[\frac{{\bar \vartheta_1^{'}}({\bar z}-{\bar
w_1})}{{\bar \vartheta_1}({\bar z}-{\bar w_1})} - \frac{{\bar
\vartheta_1^{'}}({\bar z}-{\bar w_2})}{{\bar \vartheta_1}({\bar
z}-{\bar w_2})}+\frac{\pi (w_{12}
- {\bar w_{12}}) }{\tau_2}\right].\nonumber\\
\label{EQ:freeCorrel4}
\end{eqnarray}
Lastly
\begin{eqnarray}
&\,&\langle \partial {\hat \phi} (z) {\bar \partial} {\hat
\phi}({\bar z}) \exp(i {\hat \phi}(w_1,{\bar w_1})/2) \exp(-i {\hat
\phi}(w_2,{\bar w_2})/2) \rangle \nonumber\\
&=& -\frac{1}{4}\exp \left[-\frac{\pi (w_{12}-{\bar
w_{12}})^2}{8\tau_2}\right] \left\arrowvert\frac{\vartheta'_1(0)}
{\vartheta_1(w_{12})}\right\arrowvert^{1/2}
\left\arrowvert\frac{\vartheta_1^{'}(z-w_1)}{\vartheta_1(z-w_1)} -
\frac{\vartheta_1^{'}(z-w_2)}{\vartheta_1(z-w_2)}-\frac{\pi (w_{12}
- {\bar w_{12}}) }{\tau_2}\right\arrowvert^2\nonumber\\
&-& \frac{\pi}{\tau_2}\exp \left[-\frac{\pi (w_{12}-{\bar
w_{12}})^2}{8\tau_2}\right] \left\arrowvert\frac{\vartheta'_1(0)}
{\vartheta_1(w_{12})}\right\arrowvert^{1/2} \label{EQ:freeCorrel5}
\end{eqnarray}
The second term of this equation comes from the fact that from
Eq.~(\ref{EQ:freeProp}), $\langle \partial {\hat \phi} (z) {\bar
\partial} {\hat \phi}({\bar z}) \rangle$ is actually nonzero.

Inserting Eq.~(\ref{EQ:freeCorrel1}), (\ref{EQ:freeCorrel3}),
(\ref{EQ:freeCorrel4}), and (\ref{EQ:freeCorrel5}) into
Eq.~(\ref{EQ:a1Correl1}) gives
\begin{eqnarray}
&\,&\langle \partial \phi(z){\bar \partial} \phi({\bar z}) \exp(i
\phi(w_1,{\bar w_1})/2) \exp(-i \phi(w_2,{\bar w_2})/2)
\rangle_{{\bf
\alpha} = (1/2,1/2)}\nonumber\\
&=& -\frac{\sqrt {2 \tau_2}}{Z_0}
\left\arrowvert\frac{\vartheta'_1(0)}
{\vartheta_1(w_{12})}\right\arrowvert^{1/2} \left\arrowvert
\vartheta'_1(w_{12}/2) + \frac{1}{2}\vartheta_1(w_{12}/2)
\left(\frac{\vartheta_1^{'}(z-w_1)}{\vartheta_1(z-w_1)} -
\frac{\vartheta_1^{'}(z-w_2)}{\vartheta_1(z-w_2)}\right)
\right\arrowvert^2.\nonumber\\
\label{EQ:BCorrelResult1}
\end{eqnarray}
Similarly,
\begin{eqnarray}
&\,&\langle \partial \phi(z){\bar \partial} \phi({\bar z}) \exp(-i
\phi(w_1,{\bar w_1})/2) \exp(i \phi(w_2,{\bar w_2})/2) \rangle_{{\bf
\alpha} = (1/2,1/2)}\nonumber\\
&=& -\frac{\sqrt {2 \tau_2}}{Z_0}
\left\arrowvert\frac{\vartheta'_1(0)}
{\vartheta_1(w_{12})}\right\arrowvert^{1/2} \left\arrowvert
\vartheta'_1(-w_{12}/2) - \frac{1}{2}\vartheta_1(-w_{12}/2)
\left(\frac{\vartheta_1^{'}(z-w_1)}{\vartheta_1(z-w_1)} -
\frac{\vartheta_1^{'}(z-w_2)}{\vartheta_1(z-w_2)}\right)
\right\arrowvert^2.\nonumber\\
\label{EQ:BCorrelResult2}
\end{eqnarray}

Eq.~(\ref{EQ:BCorrelResult1}) and (\ref{EQ:BCorrelResult2}),
together with Eq.~(\ref{EQ:chiralIsing}) and
(\ref{EQ:IsingBosonization}) leads to the following result stated in
Sec.III B:
\begin{eqnarray}
\langle \psi(z) \sigma(w_1) \sigma(w_2) \rangle_{\alpha = (1/2,1/2)}
&\propto&
\left[\frac{1}{\vartheta_1(w_{12})}\right]^{1/8}\nonumber\\
&\times& \left[\vartheta_1^{'}(w_{12}/2) +
\frac{1}{2}\vartheta_1(w_{12}/2)\left(\frac{\vartheta_1^{'}(z-
w_1)}{\vartheta_1(z-w_1)} -
\frac{\vartheta_1^{'}(z-w_2)}{\vartheta_1(z-w_2)}\right)\right]^{1/
2}.\nonumber\\
\end{eqnarray}

\end{document}